\documentclass[twocolumn]{aastex631}
\usepackage{CJK}
\newcommand{\sn}{SN\,2022joj}

\newcommand{\tfl}{$t_\mathrm{fl}$}
\newcommand{\Mch}{$M_\mathrm{Ch}$}
\newcommand{\kms}{$\mathrm{km}\,\mathrm{s}^{-1}$}

\shorttitle{\sn}
\shortauthors{Liu et al.}
\graphicspath{{./}{figures/}}

\begin{document}
\begin{CJK*}{UTF8}{gbsn}

\title{SN\,2022joj: A Peculiar Type Ia Supernova Possibly Driven by an Asymmetric Helium-shell Double Detonation}

%main contributors
\author[0000-0002-7866-4531]{Chang~Liu (刘畅)}
\affil{Department of Physics and Astronomy, Northwestern University, 2145 Sheridan Rd, Evanston, IL 60208, USA}
\affil{Center for Interdisciplinary Exploration and Research in Astrophysics (CIERA), Northwestern University, 1800 Sherman Ave, Evanston, IL 60201, USA}

\author[0000-0001-9515-478X]{Adam~A.~Miller}
\affil{Department of Physics and Astronomy, Northwestern University, 2145 Sheridan Rd, Evanston, IL 60208, USA}
\affil{Center for Interdisciplinary Exploration and Research in Astrophysics (CIERA), Northwestern University, 1800 Sherman Ave, Evanston, IL 60201, USA}

\author[0000-0002-1184-0692]{Samuel~J.~Boos}
\affil{Department of Physics \& Astronomy, University of Alabama, Tuscaloosa, AL, USA}

\author[0000-0002-9632-6106]{Ken~J.~Shen}
\affil{Department of Astronomy and Theoretical Astrophysics Center, University of California, Berkeley, CA 94720-3411, USA}

\author[0000-0002-9538-5948]{Dean~M.~Townsley}
\affil{Department of Physics \& Astronomy, University of Alabama, Tuscaloosa, AL, USA}

\author[0000-0001-6797-1889]{Steve~Schulze}
\affil{The Oskar Klein Centre, Department of Physics, Stockholm University, Albanova University Center, SE-106 91 Stockholm, Sweden}

\author[0000-0003-3393-9383]{Luke~Harvey}
\affil{School of Physics, Trinity College Dublin, The University of Dublin, Dublin 2, Ireland}

\author[0000-0002-9770-3508]{Kate~Maguire}
\affil{School of Physics, Trinity College Dublin, The University of Dublin, Dublin 2, Ireland}

\author[0000-0001-5975-290X]{Joel~Johansson}
\affil{The Oskar Klein Centre, Department of Physics, Stockholm University, Albanova University Center, SE-106 91 Stockholm, Sweden}

%photometry & spectra providers
\author[0000-0001-5955-2502]{Thomas~G.~Brink}
\affil{Department of Astronomy, University of California, Berkeley, CA 94720-3411, USA}

\author[0000-0003-0126-3999]{Umut~Burgaz}
\affil{School of Physics, Trinity College Dublin, The University of Dublin, Dublin 2, Ireland}

\author[0000-0001-9494-179X]{Georgios~Dimitriadis}
\affil{School of Physics, Trinity College Dublin, The University of Dublin, Dublin 2, Ireland}

\author[0000-0003-3460-0103]{Alexei~V.~Filippenko}
\affil{Department of Astronomy, University of California, Berkeley, CA 94720-3411, USA}

\author[0000-0002-3841-380X]{Saarah~Hall}
\affil{Department of Physics and Astronomy, Northwestern University, 2145 Sheridan Rd, Evanston, IL 60208, USA}
\affil{Center for Interdisciplinary Exploration and Research in Astrophysics (CIERA), Northwestern University, 1800 Sherman Ave, Evanston, IL 60201, USA}

\author[0000-0002-0129-806X]{K-Ryan~Hinds}
\affil{Astrophysics Research Institute, Liverpool John Moores University, Liverpool Science Park, 146 Brownlow Hill, Liverpool L3 5RF, UK}

\author[0000-0002-8732-6980]{Andrew~Hoffman}
\affil{Institute for Astronomy, University of Hawai\'i at Manoa, 2680 Woodlawn Dr., Honolulu, HI 96822, USA}

\author[0000-0003-2758-159X]{Viraj~Karambelkar}
\affil{Division of Physics, Mathematics and Astronomy, California Institute of Technology, Pasadena, CA 91125, USA}

\author[0000-0002-5740-7747]{Charles~D.~Kilpatrick}
\affil{Center for Interdisciplinary Exploration and Research in Astrophysics (CIERA), Northwestern University, 1800 Sherman Ave, Evanston, IL 60201, USA}

\author[0000-0002-0129-806X]{Daniel~Perley}
\affil{Astrophysics Research Institute, Liverpool John Moores University, Liverpool Science Park, 146 Brownlow Hill, Liverpool L3 5RF, UK}

\author[0009-0009-7665-6827]{Neil~Pichay}
\affil{Department of Astronomy, University of California, Berkeley, CA 94720-3411, USA}

\author[0000-0001-8023-4912]{Huei~Sears}
\affil{Department of Physics and Astronomy, Northwestern University, 2145 Sheridan Rd, Evanston, IL 60208, USA}
\affil{Center for Interdisciplinary Exploration and Research in Astrophysics (CIERA), Northwestern University, 1800 Sherman Ave, Evanston, IL 60201, USA}

\author[0000-0003-1546-6615]{Jesper~Sollerman}
\affiliation{The Oskar Klein Centre, Department of Astronomy, Stockholm University, Albanova University Center, SE-106 91 Stockholm, Sweden}

\author[0000-0003-2434-0387]{Robert~Stein}
\affil{Division of Physics, Mathematics and Astronomy, California Institute of Technology, Pasadena, CA 91125, USA}

\author[0000-0001-9834-3439]{Jacco~H.~Terwel}
\affil{School of Physics, Trinity College Dublin, The University of Dublin, Dublin 2, Ireland}
\affil{Isaac Newton Group (ING), Apt. de correos 321, E-38700, Santa Cruz de La Palma, Canary Islands, Spain}

\author[0000-0002-2636-6508]{WeiKang~Zheng}
\affil{Department of Astronomy, University of California, Berkeley, CA 94720-3411, USA}

%builders
\author[0000-0002-3168-0139]{Matthew~J.~Graham}
\affil{Division of Physics, Mathematics and Astronomy, California Institute of Technology, Pasadena, CA 91125, USA}

\author[0000-0002-5619-4938]{Mansi~M.~Kasliwal}
\affil{Division of Physics, Mathematics and Astronomy, California Institute of Technology, Pasadena, CA 91125, USA}

\author[0000-0003-0629-5746]{Leander~Lacroix}
\affil{LPNHE, CNRS/IN2P3, Sorbonne Universit\'e, Universit\'e Paris-Cit\'e, Laboratoire de Physique Nucl\'eaire et de Hautes \'Energies, 75005 Paris, France}
\affil{The Oskar Klein Centre, Department of Physics, Stockholm University, Albanova University Center, SE-106 91 Stockholm, Sweden}

\author[0000-0003-1227-3738]{Josiah~Purdum}
\affiliation{Caltech Optical Observatories, California Institute of Technology, Pasadena, CA 91125, USA}

\author[0000-0001-7648-4142]{Benjamin~Rusholme}
\affil{IPAC, California Institute of Technology, 1200 E. California Blvd, Pasadena, CA 91125, USA}

\author[0000-0002-9998-6732]{Avery~Wold}
\affil{IPAC, California Institute of Technology, 1200 E. California Blvd, Pasadena, CA 91125, USA}

\begin{abstract} 
We present observations of SN\,2022joj, a peculiar Type Ia supernova (SN\,Ia) discovered by the Zwicky Transient Facility (ZTF). SN\,2022joj exhibits an unusually red $g_\mathrm{ZTF}-r_\mathrm{ZTF}$ color at early times and a rapid blueward evolution afterward. Around maximum brightness, SN\,2022joj shows a high luminosity ($M_{g_\mathrm{ZTF},\mathrm{max}}\simeq-19.7$\,mag), a blue broadband color ($g_\mathrm{ZTF}-r_\mathrm{ZTF}\simeq-0.2$\,mag), and shallow \ion{Si}{2} absorption lines, consistent with those of overluminous, SN\,1991T-like events. The maximum-light spectrum also shows prominent absorption around 4200\,\r{A}, which resembles the \ion{Ti}{2} features in subluminous, SN\,1991bg-like events. Despite the blue optical-band colors, SN\,2022joj exhibits extremely red ultraviolet minus optical colors at maximum luminosity {($u-v\simeq0.6$\,mag and $uvw1 - v\simeq2.5$\,mag)}, suggesting a suppression of flux at $\sim$2500--4000\,\r{A}. Strong \ion{C}{2} lines are also detected at peak. We show that these unusual spectroscopic properties are broadly consistent with the helium-shell double detonation of a sub-Chandrasekhar mass ($M\simeq1\,\mathrm{M_\odot}$) carbon/oxygen (C/O) white dwarf (WD) from a relatively massive helium shell ($M_s\simeq0.04$--$0.1\,\mathrm{M_\odot}$), if observed along a line of sight roughly opposite to where the shell initially detonates. None of the existing models could quantitatively explain all the peculiarities observed in SN\,2022joj. The low flux ratio of [\ion{Ni}{2}] $\lambda$7378 to [\ion{Fe}{2}] $\lambda$7155 emission in the late-time nebular spectra indicates a low yield of stable Ni isotopes, favoring a sub-Chandrasekhar mass progenitor. The significant blueshift measured in the [\ion{Fe}{2}] $\lambda$7155 line is also consistent with an asymmetric chemical distribution in the ejecta, as is predicted in double-detonation models.
\end{abstract}


\section{Introduction} \label{sec:intro}
Type Ia supernovae (SNe\,Ia) come from thermonuclear explosions of carbon/oxygen (C/O) white dwarfs (WDs) in binary systems. While there is broad consensus about this fact, specifics about the binary companion and the conditions that spark ignition remain uncertain \citep[e.g.,][for reviews]{Maoz_2014, Liu_2023}. Multiple explosion channels have been proposed, though none of them can fully explain the diversity in the SN\,Ia population.

Recent attention has been focused on the helium-shell double-detonation scenario as a potential explanation for some normal SNe\,Ia \citep[e.g.,][]{polin_observational_2019,Shen_2D_2021} as well as a growing subclass of peculiar, red SNe\,Ia \citep[e.g.,][]{jiang_16jhr_2017, de_18byg_2019,Liu_20jgb_2023}. In a double detonation, the detonation moving along the base of a helium shell (accreted from a helium-rich companion) atop the primary WD sends an oblique shock wave inward \citep[e.g.,][]{Fink_DD_2010}, which eventually converges somewhere within the core, triggers a secondary detonation, and inevitably explodes the entire WD \citep{Nomoto_1982a,Nomoto_1982b,Woosley_1986,Livne_1990,Woosley_1994,Livne_1995}. This mechanism can dynamically ignite WDs well below the Chandrasekhar mass (\Mch; $\sim$1.4\,$\mathrm{M_\odot}$). It has been proposed that a substantial fraction of SNe\,Ia could result from double detonations based on observations of (i) the intrinsic event rate and delay-time distribution (DTD) of the SN\,Ia population \citep{Ruiter_2011,Ruiter_2014}; (ii) the nucleosynthetic yields of SNe\,Ia as measured in their late-time spectra \citep{Maguire_2018,Flors_2020}; (iii) the chemical-enrichment history of various galaxies \citep{Kirby_2019,de_los_reyes_manganese_2020,Sanders_2021,Eitner_2022}; and (iv) the hypervelocity Galactic WDs, which are likely surviving donors from double-degenerate binaries where the primary WD exploded in a double detonation \citep{Shen_2018,El-Badry_2023}.

The remarkable observational properties for SNe\,Ia from double detonations are mostly associated with the helium shell. Shortly after the shell detonation, the decay of the radioactive species synthesized during the helium burning may power a flux excess in the early-time light curves \citep{Woosley_1994,Fink_DD_2010,Kromer_DD_2010}. Afterward, the iron-group elements (IGEs) in the helium-shell ashes may provide significant line blanketing blueward of $\sim$5000\,\r{A} \citep{Kromer_DD_2010}, efficiently suppressing flux in the blue optical. In general, progenitors with a thin helium shell would show minimal detectable effects from the shell detonation, and reproduce ``normal'' \citep[e.g., that of SN\,2011fe;][]{Nugent_11fe_2011} luminosity and spectroscopic properties around maximum luminosity \citep[e.g.,][]{polin_observational_2019, Townsley_2019,Magee_2021,Shen_2D_2021}. Normal SNe\,Ia with a red flux excess shortly after the explosion may be associated with this scenario \citep[e.g., SN\,2018aoz;][]{Ni_2022}. Meanwhile, objects involving a more massive helium shell exhibit peculiarities, such as a strong flash at early times and an extremely red color around maximum luminosity \citep{polin_observational_2019}. Several peculiar SNe\,Ia have been interpreted as double-detonation SNe, including SN\,2016jhr \citep{jiang_16jhr_2017}, SN\,2018byg \citep{de_18byg_2019}, OGLE-2013-SN-079 \citep[][interpreted as either a pure helium-shell detonation or a double detonation]{Inserra_OGLE13_079_2015}, SN\,2016hnk (\citealp{jacobson-galan_16hnk_2020,de_Ca_rich_2020}; but see \citealp{galbany_16hnk_2019} for an alternative interpretation), SN\,2019ofm \citep{de_Ca_rich_2020}, SN\,2016dsg \citep{Dong_16dsg_2022}, SN\,2020jgb \citep{Liu_20jgb_2023}, and SN\,2019eix \citep{Gonzalez_19eix_2023}.

Multidimensional considerations are especially important for double detonations because the explosion of the C/O core is triggered off-center, and, as a result, all the observables (e.g., luminosities, colors, and absorption line features) are subject to viewing-angle effects \citep{Fink_DD_2010,Shen_2D_2021}. Asymmetries in the chemical distribution of the SN\,Ia ejecta have been invoked to explain SN spectropolarimetric measurements \citep[e.g.,][see \citealp{Wang_2008} for a review]{Wang_2003,Kasen_2003, Patat_2012} and the kinematics of Ni and Fe in the innermost ejecta \citep{Motohara_2006,Maeda_2010b,Maeda_2010,Maguire_2018,Li_2021}. To accurately infer the progenitor of a double-detonation SN, one needs to compare the observations with multidimensional models.

In this paper, we present observations of a peculiar SN\,Ia, \sn, which shows a remarkable color evolution, starting with red optical colors that quickly evolve to the blue as the SN rises to maximum luminosity. Its photometric and spectroscopic features are qualitatively consistent with that of a double-detonation SN. In Section~\ref{sec:obs}, we summarize the observations of \sn, which are analyzed in Section~\ref{sec:analysis}, where we show its peculiarities in various aspects. In Section~\ref{sec:model}, we discuss existing scenarios that can lead to a red color in the early light curves of an SN\,Ia, of which the helium-shell double-detonation scenario is the most reasonable explanation. We also show that multidimensional effects must be taken into account to explain the spectroscopic peculiarities. 
The indication of a sub-\Mch\ progenitor and an asymmetric explosion is supported by the late-time spectra of \sn, which we describe in Section~\ref{sec:disc_nebular}. In Section~\ref{sec:disc_C_II}, we discuss possible origins of the carbon features in \sn\ at maximum brightness.
We draw our conclusions in Section~\ref{sec:conclusion}. {After the submission of this paper, a separate study of \sn\ by \citet{Padilla_Gonzalez_2023} was posted on the arXiv, drawing similar broad conclusions.}

Along with this paper, we have released the data utilized in this study and the software used for data analysis and visualization. They are available online at {Zenodo under an open-source 
Creative Commons Attribution license: \dataset[10.5281/zenodo.8331024]{\doi{10.5281/zenodo.8331024}} and our GitHub repository, \url{https://github.com/slowdivePTG/SN2022joj}.}

\section{Observations} \label{sec:obs}
\subsection{Discovery \& Classification}
\sn\ was discovered by the Zwicky Transient Facility \citep[ZTF;][]{Bellm_ZTF_2019a,Graham_ZTF_2019,Dekany_ZTF_2020} on 2022 May 08.298 (UTC dates are used throughout the paper; MJD 59707.298) with the 48\,inch Samuel Oschin Telescope (P48) at Palomar Observatory, via the ZOGY image-differencing algorithm \citep{Zackay_imagesub_2016}, which is utilized by the automated ZTF discovery pipeline \citep{Masci_ZTF_2019}. It was first detected with $r_\mathrm{ZTF}=19.13\pm0.06$\,mag at $\alpha_\mathrm{J2000}=14^\mathrm{h}41^\mathrm{m}40\fs08$, $\delta_\mathrm{J2000}=+03\degr00'24\farcs{14}$ and announced to the public by \citet{Fremling_2022TNSTR}. A real-time alert \citep{Patterson_ZTFalert_2019} was generated as the candidate passed internal machine-learning thresholds \citep[e.g.,][]{Duev_ZTFML_2019,Mahabal_ZTFML_2019}, and the internal designation ZTF22aajijjf was assigned. The follow-up observations of the SN was coordinated using the Fritz Marshal \citep{van_der_Walt_skyportal_2019,Coughlin_skyportal_2023}. The last 3$\sigma$ nondetection limits the brightness to $r_\mathrm{ZTF}>21.48$\,mag on 2022 May 03.27 (MJD 59702.27; 5.03\,days before the first detection) using the ZTF forced photometry from the \texttt{ZTF Forced Photometry Survice} \citep[\texttt{ZFPS};][]{Masci_ZTFforced_2023}. \sn\ was also independently monitored by the Asteroid Terrestrial-impact Last Alert System \citep[ATLAS;][]{ATLAS_2018,ATLAS_2020}. With the forced photometry obtained from the ATLAS forced-photometry server \citep{ATLAS_forced_phot_2021},\footnote{\url{https://fallingstar-data.com/forcedphot/}} we identify the last 3$\sigma$ nondetection with ATLAS on 2022 May 04.26, 0.99\,days after the last nondetection in $r_\mathrm{ZTF}$, and put a limit of the brightness in the orange filter of $o>19.84$\,mag.

The first spectrum was obtained on 2022 May 11.288 by \citet{Newsome_2022TNSCR}, who found a best fit to a young Type I SN at redshift $z=0.03$ using the \texttt{Supernova Identification (SNID)} algorithm \citep{Blondin_SNID_2007}. In this early-time spectrum, prominent \ion{Si}{2} $\lambda$6355 and \ion{Ca}{2} infrared triplet (IRT) absorption suggest an SN\,Ia classification, but the overall spectral shape, featuring a relatively red continuum (see Figure~\ref{fig:spec_seq}), is atypical for a normal SN\,Ia at this phase. \citet{Chu_2022TNSCR} used a peak-luminosity spectrum to indisputably classify \sn\ as an SN\,Ia based on its blue color and persistent \ion{Si}{2} features.

\subsection{Host Galaxy}\label{sec:host}
The host of \sn\ is a dwarf galaxy at $\alpha_\mathrm{J2000}=14^\mathrm{h}41^\mathrm{m}40\fs04$, $\delta_\mathrm{J2000}=+03\degr00'24\farcs{53}$, cataloged in the DESI Legacy Imaging Survey \citep[LS;][]{Dey_LS_2019}, which reports 3$\sigma$ detections in $grz$ and $W_1$ (see Table~\ref{tab:host_phot}). \sn\ has a projected offset of only $0\farcs{5}\pm0\farcs{1}$ from the host (corresponding to a projected distance of $0.27\pm0.05$\,kpc at the redshift estimated below). 

In addition, the SN field was observed in $grizy$ as part of the wide survey of the Hyper Suprime-Cam Subaru Strategic Program \citep[HSC-SSP;][]{Aihara2018a}. We retrieved the stacked science-ready images from the HSC-SSP data archive using the HSC data-access tools.\footnote{\href{https://hsc-gitlab.mtk.nao.ac.jp/ssp-software/data-access-tools}{https://hsc-gitlab.mtk.nao.ac.jp/ssp-software/data-access-tools}} The photometry was extracted with the aperture-photometry tool presented by \citet{Schulze2018a}. The measurements were calibrated against a set of stars from the Pan-STARRS catalog \citep{PS1_2016}, and we applied color terms from the HSC pipeline version 8\footnote{\href{https://hsc.mtk.nao.ac.jp/pipedoc/pipedoc_8_e/colorterms.html}{https://hsc.mtk.nao.ac.jp/pipedoc/pipedoc\_8\_e/\\colorterms.html}} to correct for differences between the Pan-STARRS and HSC filters. Table \ref{tab:host_phot} summarizes all measurements.

To determine the redshift of the host, we obtained two spectra about 300\,days after the SN maximum brightness. On 2023 March 14, we took a spectrum of both the SN and the host using Binospec \citep{Binospec_2019} on the 6.5\,m MMT telescope with a total integration time of 5400\,s. We placed the slit across both the center of the galaxy and the position of the SN (Figure~\ref{fig:host_spec}). On 2023 April 26, we took another spectrum using the Low Resolution Imaging Spectrometer \citep[LRIS;][]{LRIS_1995} on the Keck I 10\,m telescope. The slit was placed at the same position angle, with the Cassegrain Atmospheric Dispersion Compensator \cite[Cass ADC;][]{{LRIS_ADC_2006}} module on. The total integration time was 3600\,s.
The LRIS spectrum has a higher signal-to-noise ratio (S/N), in which we detected a potential host emission line at 6742.4\,\r{A} (see Figure~\ref{fig:host_spec}) with a S/N\footnote{We fit the emission line with a Gaussian profile to estimate its intensity, and the S/N is defined as the intensity divided by its uncertainty.} of 3.4. We associated this feature with H$\alpha$ emission, meaning the corresponding redshift of the host galaxy is $z=0.02736\pm0.0007$, in agreement with the initial estimate, $z=0.03$, from matching the SN spectra to \texttt{SNID} templates \citep{Newsome_2022TNSCR}. In the coadded two-dimensional (2D) spectrum, the trace is dominated by the light of the SN in the nebular phase, while the center of this emission feature has an offset of $\sim$3--4 pixels from the center of the trace. The CCDs on LRIS have a scale of $0\farcs135$\,pixel$^{-1}$, so this offset corresponds to an angular offset of $\sim$$0\farcs4$--$0\farcs5$, consistent with the astrometric offset when comparing the LS detection of the host and the ZTF detection of the SN. The Binospec spectrum has a lower S/N, and we cannot identify this emission line at the same position in the 2D spectrum via visual inspection. Nevertheless, we still marginally detect an emission feature in the 1D spectrum with a S/N of 1.5 at the same wavelength. All the evidence indicates that the H$\alpha$ detection is real.

We estimate the distance modulus of \sn\ in the following way. We first use the 2M++ model \citep{Carrick2015_2M++} to estimate the peculiar velocity of the host galaxy to be $244\pm250$\,\kms. Then the peculiar velocity is combined with the recession velocity in the frame of the cosmic microwave background (CMB) $v_\mathrm{CMB}=8424$\,\kms, which yields a net Hubble recession rate of $8193\pm250$\,\kms. Using cosmological parameters $H_0 = 70\,\mathrm{km\,s^{-1}\,Mpc^{-1}}$, $\Omega_M=0.3$, and $\Omega_\Lambda=0.7$, the estimated luminosity distance to \sn\ is 119.5\,Mpc, equivalent to a distance modulus of $35.39\pm0.03$\,mag.

\begin{deluxetable}{cccc} \label{tab:host_phot}
\tabletypesize{\scriptsize}
\tablewidth{0pt}
\tablecaption{Host photometry of \sn.}
\tablehead{
    \colhead{Survey} & \colhead{Filter} & \colhead{$m$} & \colhead{$\sigma_m$}\\
    \colhead{} & \colhead{} & \colhead{(mag)} & \colhead{(mag)}
}
\startdata
HSC-SSP & PS $g$        &$ 22.01$ & $0.03$\\
HSC-SSP & PS $r$        &$ 21.63$ & $0.02$\\
HSC-SSP & PS $i$        &$ 21.54$ & $0.03$\\
HSC-SSP & PS $z$        &$ 21.37$ & $0.04$\\
HSC-SSP & PS $y$        &$ 21.22$ & $0.09$\\
LS & DECam $g$           &$ 22.05$ & $0.05$\\
LS & DECam $r$           &$ 21.65$ & $0.05$\\
LS & DECam $z$           &$ 21.39$ & $0.10$\\
LS & $W_1$           &$ 21.94$ & $0.35$
\enddata
\tablecomments{HSC-SSP -- Hyper Suprime-Cam Subaru Strategic Program; LS -- the DESI Legacy Imaging Survey; PS -- Pan-STARRS; DECam -- the Dark Energy Camera. All magnitudes are reported in the AB system \citep{Oke_1983} and are not corrected for reddening.}
\end{deluxetable}

\subsection{Optical Photometry}
\begin{figure*}
    \centering
    \includegraphics[width=\textwidth]{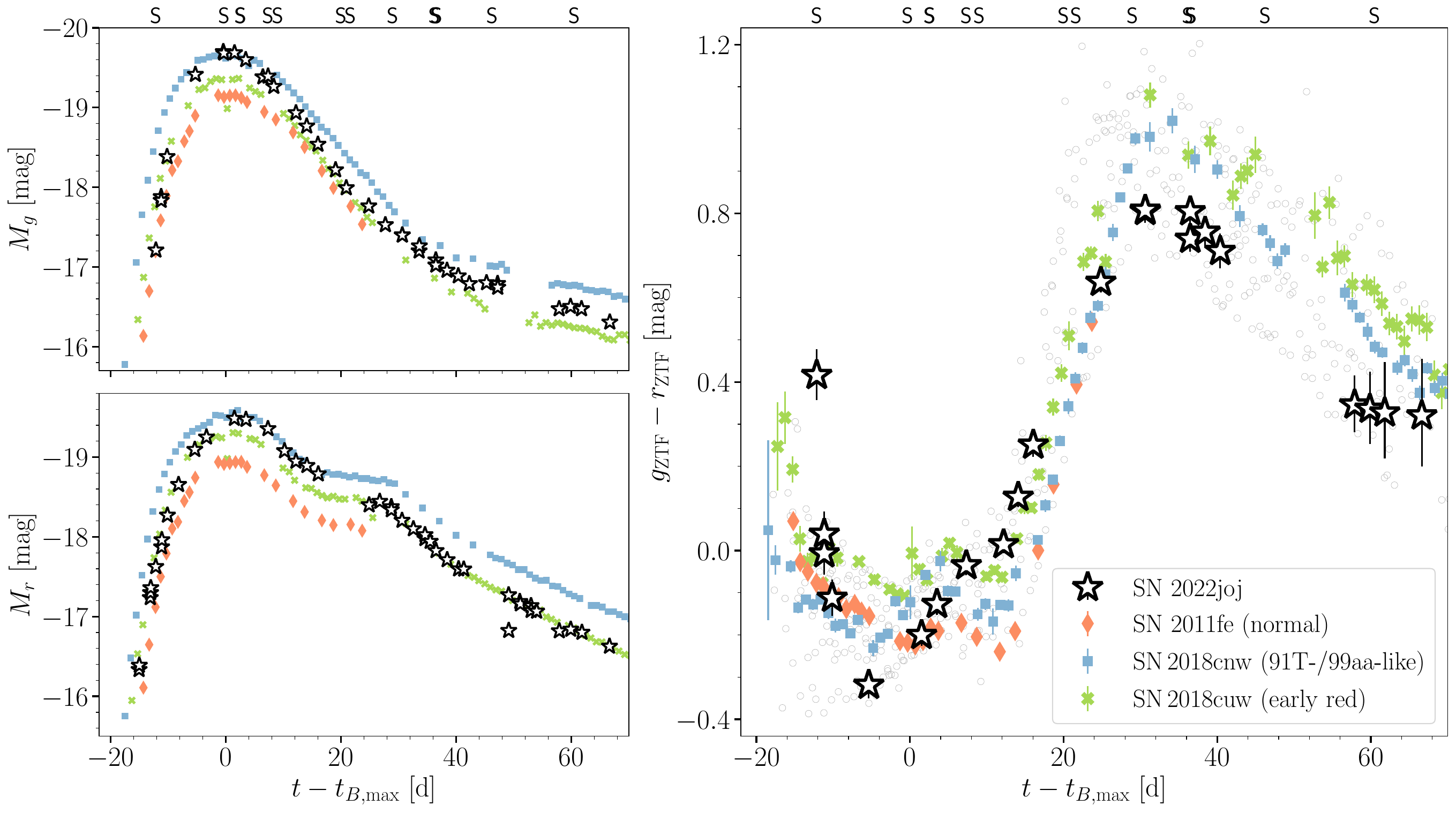}
    \caption{\sn\ (black stars) is an overluminous, fast-evolving SN\,Ia with a unique color evolution. In comparison, we show the photometric properties of SN\,2011fe \citep[normal SN\,Ia;][]{Pereira_2013}, SN\,2018cnw (91T-/99aa-like), and SN\,2018cuw (normal SN\,Ia with a red early color). \textit{Left}: multiband light curves. The upper (lower) panel shows the evolution in the $g$-band ($r$-band) absolute magnitude.
    \textit{Right}: $g_\mathrm{ZTF}-r_\mathrm{ZTF}$ color evolution. 
    The gray circles denote the color evolution of 14 nearby ($z\le0.05$) SNe\,Ia (open circles) from the ZTF sample with prompt observations within 5\,days of first light \citep{Bulla2020}. The phases {with respect to $t_{B, \mathrm{max}}$ (MJD 59722.77)} are measured in the rest frame of the host galaxy. Correction for Galactic extinction has been applied, but $K$-corrections have not been performed. The epochs of optical spectroscopy are marked with ``S'' on the top axis.}
    \label{fig:lc}
\end{figure*}

\sn\ was monitored in ZTF $gri$ bands as part of its ongoing Northern Sky Survey \citep{Bellm_ZTF_2019b}. The $i_\mathrm{ZTF}$ data do not cover the rise. We use the forced-photometry light curves from \texttt{ZFPS}, reduced using the pipeline from A. A. Miller et al. (2023, in prep.); see also \citet{Yao_2019}. We adopt a Galactic extinction of ${E(B-V)_\mathrm{MW}}=0.032$\,mag \citep{Schlafly2011}, and correct all photometry using the extinction model from \citet{Fitzpatrick1999} assuming $R_V=3.1$. We do not find any \ion{Na}{1} D absorption at the redshift of the host galaxy (we put a 3$\sigma$ upper limit in the equivalent width of \ion{Na}{1} $\mathrm{D_1}+\mathrm{D_2}$ of $<$$0.5$\,\r{A}), indicating that the extinction from the host is negligible. The blue $g_\mathrm{ZTF}-r_\mathrm{ZTF}$ color ($\sim$$-0.2$\,mag) near maximum luminosity after correcting for the Galactic extinction is also consistent with no additional reddening from the host. Therefore we assume ${E(B-V)}_\mathrm{host}=0$. The dereddened $g_\mathrm{ZTF}$ and $r_\mathrm{ZTF}$ forced-photometry light curves in absolute magnitudes are shown in Figure~\ref{fig:lc}. 
Additional observations of \sn\ were obtained in the $o$ and $c$ filters in the ATLAS survey, the $griz$ filters with the optical imaging component of the Infrared-Optical suite of instruments (IO:O) on the Liverpool Telescope, the $griBVRI$ filters and the clear filter on the 0.76\,m Katzman Automatic Imaging Telescope \citep[KAIT;][]{KAIT_2001} at Lick Observatory, and the $BVRI$ filters on the 1\,m Anna Nickel telescope at Lick. ZTF, ATLAS, LT, KAIT, and Nickel observations are reported in Table~\ref{tab:phot}. {The ZTF $gri$ magnitudes, ATLAS $o$ and $c$ magnitudes, and Sloan $griz$ magnitudes are reported in the AB system, while the $BVRI$ and clear magnitudes are reported in the Vega system.}

\subsection{Swift Ultraviolet/Optical Telescope (UVOT) Observations}
Ultraviolet (UV) observations of \sn\ were obtained using UVOT \citep{UVOT_2005} on the {\it Neil Gehrels Swift Observatory} \citep[{\it Swift};][]{Swift_2004} following a target-of-opportunity (ToO) request by E.~Padilla Gonzalez. Prior to the SN, UVOT images of the field had been obtained in the $u$, $uvw1$, and $uvw2$ filters. In each of these reference images the flux at the location of the SN is consistent with 0, and the 3$\sigma$ upper limits correspond to $\lesssim$10\% of the SN flux measured in the UV. In the optical bands, the LS photometry ($g = 22.05\pm0.05$\,mag) shows that the host galaxy contributes $\lesssim$1\% of the total observed flux. We therefore conclude that the host galaxy can be neglected when estimating the SN flux in UVOT images.

We determine the flux in the $u$, $b$, $v$, $uvw1$, $uvm2$, and $uvw2$ filters using a circular aperture with a radius of $5''$ centered at the position of the SN. To estimate the brightness of the sky background we use a coaxial annulus region with an inner/outer radius of $8''$/$15''$. The reduction is performed with the pipeline \texttt{Swift\_ToO}\footnote{\url{https://github.com/slowdivePTG/Swift\_ToO}} based on the package \texttt{HEAsoft}\footnote{\url{http://heasarc.gsfc.nasa.gov/ftools}} version 6.30.1 \citep{HEAsoft_2014}. The {magnitudes} are also reported in Table~\ref{tab:phot} {in the Vega system}.

\begin{deluxetable}{cccccc} \label{tab:phot}
    \tabletypesize{\scriptsize}
    \tablewidth{0pt}
    \tablecaption{Optical and UV Photometry of \sn.}
    \tablehead{
    \colhead{$t_\mathrm{obs}$} &
    \colhead{Filter} &
    \colhead{$m$} &
    \colhead{$\sigma_m$} &
    \colhead{Magnitude} &
    \colhead{Telescope} \\
    \colhead{(MJD)} &
    \colhead{} &
    \colhead{(mag)} &
    \colhead{(mag)} &
    \colhead{System} &
    \colhead{}
    }
    \startdata
    59707.298 & ZTF $r$ & 19.131 & 0.064 & AB & P48/ZTF \\
    59707.340 & ZTF $r$ & 19.078 & 0.043 & AB & P48/ZTF \\
    59709.295 & ZTF $r$ & 18.229 & 0.037 & AB & P48/ZTF \\
    59709.342 & ZTF $r$ & 18.167 & 0.030 & AB & P48/ZTF \\
    59709.381 & ZTF $r$ & 18.105 & 0.023 & AB & P48/ZTF \\
    \enddata
    \tablecomments{Observed magnitudes in the ZTF, ATLAS, UVOT, LT, KAIT, and Nickel passbands. Correction for Galactic extinction has not been applied.\\(This table is available in its entirety in machine readable form.)}
\end{deluxetable}

\subsection{Optical Spectroscopy}\label{sec:optical_spec}

\begin{figure}
    \centering
    \includegraphics[width=\linewidth]{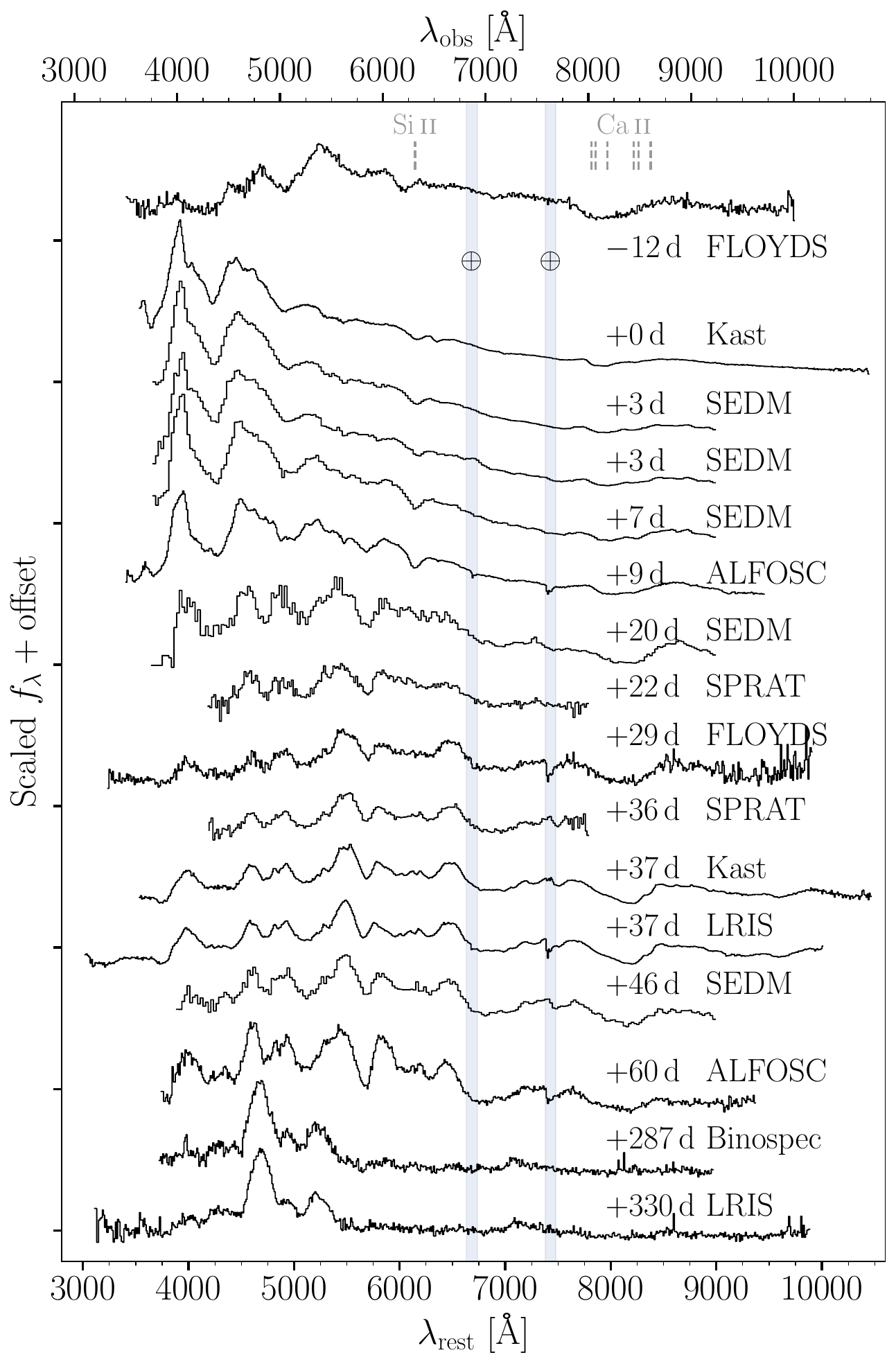}
    \caption{Optical spectral sequence of \sn\ showing the red to blue color transition as the SN rises to its maximum luminosity and the development of prominent absorption features around 4200\,\r{A} post-maximum.
    Rest-frame phase relative to the $B$-band peak and the instrument used to observe the SN are listed for each spectrum. Spectra have been corrected for ${E(B-V)_\mathrm{MW}} = 0.032$\,mag. All spectra are binned with a bin size of 10\,\AA, except for the low-resolution SEDM spectra. The corresponding wavelengths of the \ion{Si}{2} $\lambda$6355 line (with an expansion velocity of 10,000\,\kms) and the \ion{Ca}{2} IRT (with expansion velocities of both 10,000\,\kms\ and 25,000\,\kms) are marked by the vertical dashed lines. The strong optical telluric features (Fraunhofer A and B bands) are marked by the shaded region.}
    \label{fig:spec_seq}
\end{figure}

\begin{figure*}
    \centering
    \includegraphics[width=\linewidth]{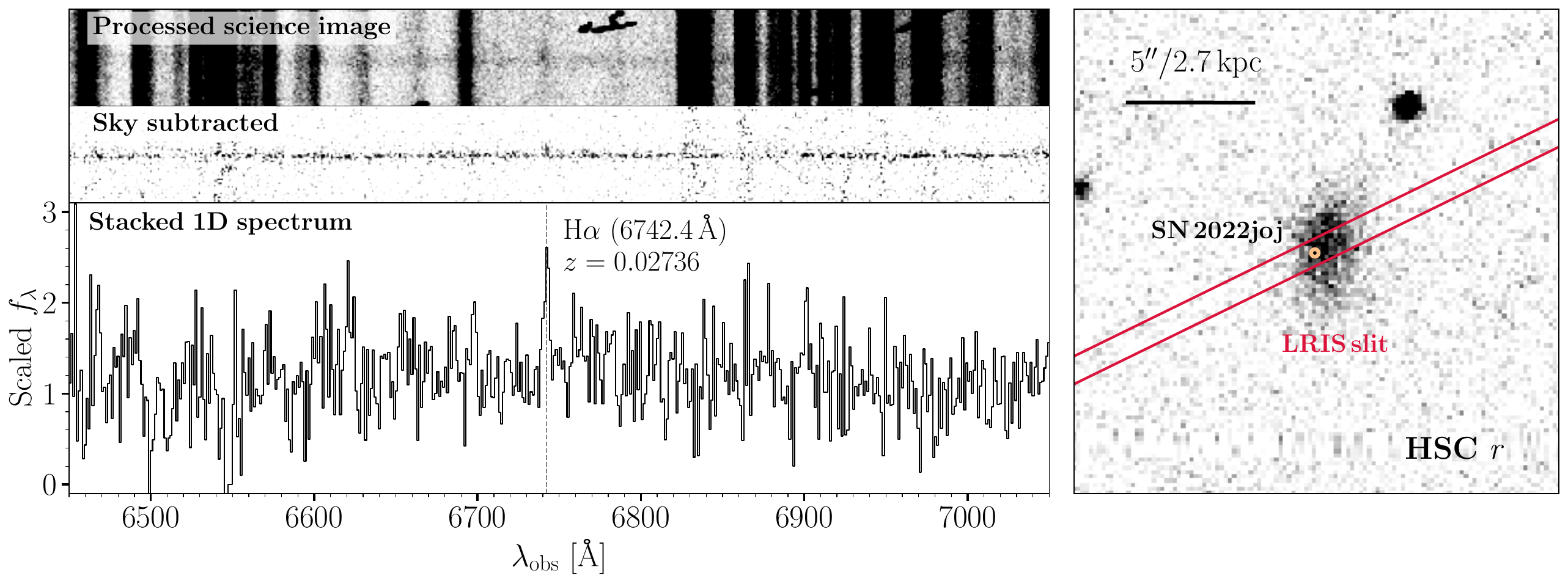}
    \caption{The LRIS spectrum reveals an H$\alpha$ emission line from the host galaxy at 6742.4\,\AA, corresponding to $z=0.02736$. \textit{Left:} the H$\alpha$ emission in the observed 2D and 1D spectra. This emission line sits in the region free of strong night-sky lines and is unlikely due to bad sky subtraction.
    \textit{Right:} image of the host galaxy and the position of \sn, with the orientation of the LRIS slit overplotted. 
    }
    \label{fig:host_spec}
\end{figure*}

We obtained a series of optical spectra of \sn\ using the Spectral Energy Distribution Machine \citep[SEDM;][]{SEDM_2018} on the automated 60\,inch telescope \citep[P60;][]{P60_2006} at Palomar Observatory, the Kast double spectrograph \citep{miller1994kast} on the Shane 3\,m telescope at Lick Observatory, the Andalucia Faint Object Spectrograph and Camera (ALFOSC)\footnote{\url{http://www.not.iac.es/instruments/alfosc/}} installed at the 2.56\,m Nordic Optical Telescope (NOT), the SPectrograph for the Rapid Acquisition of Transients \citep[SPRAT;][]{SPRAT_2014} on the 2\,m Liverpool Telescope \citep[LT;][]{LT_2004} under program PL22A13 (PI: Dimitriadis), the FLOYDS spectrograph\footnote{\url{https://lco.global/observatory/instruments/floyds/}} on the 2\,m Faulkes Telescope South (FTS) at Siding Spring as part of the Las Cumbres Observatory (LCO) \citep{LCOGT_2013}, Binospec on the 6.5\,m MMT telescope, and LRIS on the Keck I 10\,m telescope. The SEDM spectra were reduced using the custom \texttt{pysedm} software package \citep{Rigault_pysedm_2019}. The Shane/Kast spectra, obtained with the slit near the parallactic angle to minimize differential slit losses \citep{Shane_1982}, were reduced following standard techniques for CCD processing and spectrum extraction \citep{Silverman_UCBIa_2012} utilizing \texttt{IRAF} \citep{IRAF_1986} routines and custom Python and IDL codes.\footnote{\url{https://github.com/ishivvers/TheKastShiv}} The NOT/ALFOSC, Keck I/LRIS, and MMT/Binospec spectra were reduced using the \texttt{PypeIt} package \citep{pypeit:joss_pub}. The LT/SPRAT spectra were reduced with a dedicated pipeline\footnote{\url{https://github.com/LivTel/sprat\_l2\_pipeline}} for bias subtraction, flat fielding, derivation of the wavelength solution and flux calibration, with additional \texttt{IRAF/PyRAF}\footnote{IRAF is distributed by the National Optical Astronomy Observatory, which is operated by the Association of Universities for Research in Astronomy (AURA) under a cooperative agreement with the U.S. National Science Foundation (NSF).} routines for proper extraction of the spectra. The FTS/FLOYDS spectrum was reduced using the FLOYDS pipeline.\footnote{\url{https://lco.global/documentation/data/floyds-pipeline/}} We also attempted to obtain a near-infrared spectrum with the Triple Spectrograph (TSpec)\footnote{\url{https://sites.astro.caltech.edu/palomar/observer/200inchResources/tspecspecs.html}} installed at the 200\,inch Hale telescope \citep[P200;][]{P200_1982} at Palomar observatory, but the observations are mostly characterized by low S/N and a few broad undulations with no immediately identifiable lines. Thus, the TSpec data were excluded from our analysis. Details of the spectroscopic observations are listed in Table~\ref{tab:spec}. The resulting spectral sequence is shown in Figure~\ref{fig:spec_seq}. All of the spectra listed in Table~\ref{tab:spec} will be available on WISeREP \citep{wiserep_2012}.

We also include the spectrum uploaded to the Transient Name Server (TNS) by \citet{Newsome_2022TNSCR} in our analysis, which was obtained using the FLOYDS spectrograph on the 2\,m Faulkes Telescope North (FTN) at Haleakala.

\begin{deluxetable}{lrccc} \label{tab:spec}
\tabletypesize{\scriptsize}
\tablewidth{0pt}
\tablecaption{Spectroscopic observations of \sn\label{tab:spectra}.}
\tablehead{
\colhead{$t_\mathrm{obs}$} &
\colhead{Phase} &
\colhead{Telescope/} &
\colhead{$R$} &
\colhead{Range} \\
\colhead{(MJD)} &
\colhead{(days)} &
\colhead{Instrument} &
\colhead{$(\lambda/\Delta\lambda)$} &
\colhead{(\AA)}
}
\startdata
59,710.29 &  $-$12.1 & FTN/FLOYDS-N & 550 & 3500--10000\\
59,722.43 &  $-$0.3  & Shane/Kast & 750 & 3630--10730 \\
59,725.34 &  $+$2.6  & P60/SEDM & 100 & 3770--9220 \\
59,725.43 &  $+$2.6  & P60/SEDM & 100 & 3770--9220 \\
59,730.27 &  $+$7.4  & P60/SEDM & 100 & 3770--9220 \\
59,732.02 &  $+$9.0  & NOT/ALFOSC & 360 & 3500--9700 \\
59,743.28 & $+$20.0  & P60/SEDM &
 100 & 3770--9220 \\
59,744.24 & $+$20.9  & P200/TSpec & 2500 & 10000--24630 \\
59,744.96 & $+$21.6  & LT/SPRAT & 350 & 4020--7990 \\
59,752.50 & $+$29.0  & FTS/FLOYDS-S & 550 & 3500--10000 \\
59,759.92 & $+$36.2  & LT/SPRAT & 350 & 4020--7990 \\
59,760.28 & $+$36.5  & Shane/Kast & 750 & 3630--10750 \\
59,760.37 & $+$36.7  & Keck I/LRIS & 1100 & 3100--10280 \\
59,770.25 & $+$46.3  & P60/SEDM & 100 & 3770--9220 \\
59,784.89 & $+$60.5  & NOT/ALFOSC & 280 & 3850--9620 \\
60,017.42 & $+$286.9 & MMT/Binospec & 1340 & 3830--9210 \\
60,061.56 & $+$329.8 & Keck I/LRIS & 1100 & 3200--10150 \\
\enddata
\tablecomments{Phase is measured relative to the $B$-band peak in the rest frame of the host galaxy. The resolution $R$ is reported for the central region of the spectrum.}
\end{deluxetable}

\section{Analysis} \label{sec:analysis}
\subsection{Early Light Curves and the First Light}
To estimate the time of first light ($t_\mathrm{fl}$), we assume an initial power-law rise in the broad-band flux $f(t)$,
$$
f(t) = A (t-t_\mathrm{fl})^\alpha,
$$
where $A$ is a constant and $\alpha$ is the power-law index. We only include the forced-photometry light curve with flux $\le$40\% of peak luminosity \citep{Miller_ZTF_2020} in $r_\mathrm{ZTF}$ and ATLAS $o$, in which observations were conducted on more than three nights between $-20$\,days and $-10$\,days. Light curves in other bands are excluded because the coverage is significantly worse at this phase (see Section~\ref{sec:analysis_phot}). We assume that the \tfl\ is the same in both bands, then estimate $\alpha$ and $A$ in each band with a Bayesian approach. We adopt flat priors for $t_\mathrm{fl}$ and $\log A$, and a normal prior for each $\alpha$ centered at 2 (the fireball model) with a standard deviation of 1. We sample their posterior distributions with Markov Chain Monte Carlo (MCMC) using the package \texttt{PyMC} \citep{pymc_2016}. In addition, we run another model with a fixed $\alpha=2$. The estimated model parameters are listed in Table~\ref{tab:basics}. We find that both light curves are consistent with a power-law rise since MJD $59703.16^{+0.70}_{-0.58}$.
This estimate is consistent with the $\alpha=2$ fireball model. When fixing $\alpha=2$, the model also fits the light curve well, but the estimated $t_\mathrm{fl}$ is $\sim$0.5\,day later (MJD $59703.66_{-0.11}^{+0.10}$). We do not find any correlated residuals as evidence for a flux excess after $\sim$4\,days since $t_\mathrm{fl}$, although a flux excess before the first detection could not be ruled out.

\subsection{Photometric Properties} \label{sec:analysis_phot}
The basic photometric properties of \sn\ are listed in Table~\ref{tab:basics}. The times of the maximum luminosity and the corresponding magnitudes in the ZTF $gr$ bands and the KAIT/Nickel $BVRI$ bands are estimated using a fourth-order polynomial fit. We do not include the maximum $i_\mathrm{ZTF}$-band properties, which are relatively uncertain owing to the low cadence in $i_\mathrm{ZTF}$ around peak. 

\sn\ shows a few peculiar photometric features compared to normal SNe\,Ia. In Figure~\ref{fig:lc}, we compare the $g_\mathrm{ZTF}$ and $r_\mathrm{ZTF}$ light curves and the $g_\mathrm{ZTF} - r_\mathrm{ZTF}$ color evolution of \sn\ with those of the well-observed normal SN\,Ia, SN\,2011fe,\footnote{We show the synthetic photometry in $g_\mathrm{ZTF}$ and $r_\mathrm{ZTF}$ calculated using the spectrophotometric sequence from \citet{Pereira_2013}.} as well as SN\,2018cnw (ZTF18abauprj) and SN\,2018cuw (ZTF18abcflnz) from a sample of SNe\,Ia with prompt observations within 5\,days of first light by ZTF \citep{Yao_2019,Bulla2020}. 
SN\,2018cnw is slightly overluminous at peak, and belongs to either the SN\,1999aa-like \citep[99aa-like;][]{Garavini_99aa_2004} or SN\,1991T-like \citep[91T-like;][]{Filippenko_91T_1992} subclass of SNe\,Ia, while SN\,2018cuw is a normal SN\,Ia with a red $g_\mathrm{ZTF} - r_\mathrm{ZTF}$ color comparable to that of \sn\ $\sim$15\,days prior to peak. 

%light curve shapes
Around maximum brightness, \sn\ is overluminous, comparable to SN\,2018cnw, and $\sim$0.5\,mag brighter than SN\,2011fe in both $g_\mathrm{ZTF}$ and $r_\mathrm{ZTF}$. But \sn\ clearly stands out owing to its fast evolution in $g_\mathrm{ZTF}$. While \sn\ and SN\,2018cnw show a similar maximum brightness in $g_\mathrm{ZTF}$, upon the first detection of \sn\ in $g_\mathrm{ZTF}$ at $\sim$$-12$\,days, its corresponding absolute magnitude ($-17.2$\,mag) is $\sim$0.8\,mag fainter than that of SN\,2018cnw at a similar phase. This means on average, \sn\ rises faster than SN\,2018cnw by $\sim$$0.06\,\mathrm{mag\,day^{-1}}$ in $g_\mathrm{ZTF}$ during that period of time. On the decline, the $\Delta m_{15}(g_\mathrm{ZTF})$ of \sn\ is $1.03\pm0.03$\,mag, which is significantly greater than that of the overluminous SN\,2018cnw ($\Delta m_{15}(g_\mathrm{ZTF})=0.77$\,mag) or normal SN\,2011fe ($\Delta m_{15}(g_\mathrm{ZTF})=0.80$\,mag). The rapid decline of \sn\ is atypical for overluminous SNe\,Ia, which are usually the slowest decliners in the SN\,Ia population \citep{Phillips_1999, Taubenberger_2017}. The rapid decline is probably due to the unusual and fast-developing absorption feature near 4200\,\r{A} (see Section~\ref{sec:analysis_spec}).

%optical color evolution
The color evolution of \sn\ does not match that of normal SNe\,Ia, as shown by the trail traced by \sn\ in the right panel of Figure~\ref{fig:lc}. We overplot all the SNe from the ZTF early SN\,Ia sample \citep{Bulla2020} with $z\le0.05$. They are corrected for Galactic extinction, but $K$-corrections have not been performed for consistency. Given the peculiar nature of \sn, we cannot use models trained on normal SNe\,Ia to reliably estimate its $K$-correction. Nevertheless, given its relatively low redshift ($z\lesssim0.03$), the $K$-corrections are not expected to be large ($K(g_\mathrm{ZTF}-r_\mathrm{ZTF})\simeq-0.05$\,mag around maximum, estimated using the Kast spectrum at $-0.3$\,days). For the same reason we only include the SNe\,Ia with the lowest redshift from the sample of \citet{Bulla2020}. 
\sn\ is remarkably red ($g_\mathrm{ZTF} - r_\mathrm{ZTF}\simeq0.4$\,mag) {at $\sim$$-12$\,days ($\sim$7\,days after $t_\mathrm{fl}$)}, and is clearly an outlier compared to the normal SN\,Ia sample.\footnote{There is one point close to the first detection of \sn\ in the color evolution diagram, which belongs to SN\,2018dhw (ZTF18abfhryc). This single $g_\mathrm{ZTF}-r_\mathrm{ZTF}$ measurement has an uncertainty of $\sim$0.1\,mag and is 2$\sigma$ redder than measurements made the nights before and after.} 
During the ensuing week, \sn\ quickly evolves to the blue, and is among the bluest objects in the sample {at $\sim$$-5$\,days} ({$\sim$14\,days after $t_\mathrm{fl}$;} $g_\mathrm{ZTF} - r_\mathrm{ZTF}\simeq -0.3$\,mag). SN\,2018cuw has a comparable $g_\mathrm{ZTF} - r_\mathrm{ZTF}$ color at early times, but the blueward evolution of SN\,2018cuw is slower than that of \sn. {No later than \sn\ reaches its peak luminosity}, \sn\ starts to evolve redward. While other SNe\,Ia show qualitatively similar redward evolution, this usually happens much later ({$\sim$$+10$\,days}). 
When $g_\mathrm{ZTF} - r_\mathrm{ZTF}$ reaches maximum ($\sim$0.8\,mag) {at $\sim$$+30$\,days}, \sn\ is again bluer than most of the SNe\,Ia in the ZTF sample. 
Eventually as \sn\ steps into the transitional phase, its color evolution follows the Lira law\footnote{The original Lira law was discovered in the $B-V$ color, but in the $g_\mathrm{ZTF}-r_\mathrm{ZTF}$ color we see a similar trend.} \citep{Lira_1996,Phillips_1999} and shows no significant difference from that of the SNe\,Ia in the ZTF sample.
The $B-V$ color evolves in the similar way as the $g_\mathrm{ZTF}-r_\mathrm{ZTF}$ color, which starts red ($B-V\simeq1.2$\,mag {at $-12$\,days}) and quickly turns bluer{, reaching $B-V\simeq0.0$\,mag at $\sim$$-5$\,days before turning red again.}

\begin{figure*}
    \centering
    \includegraphics[width=\linewidth]{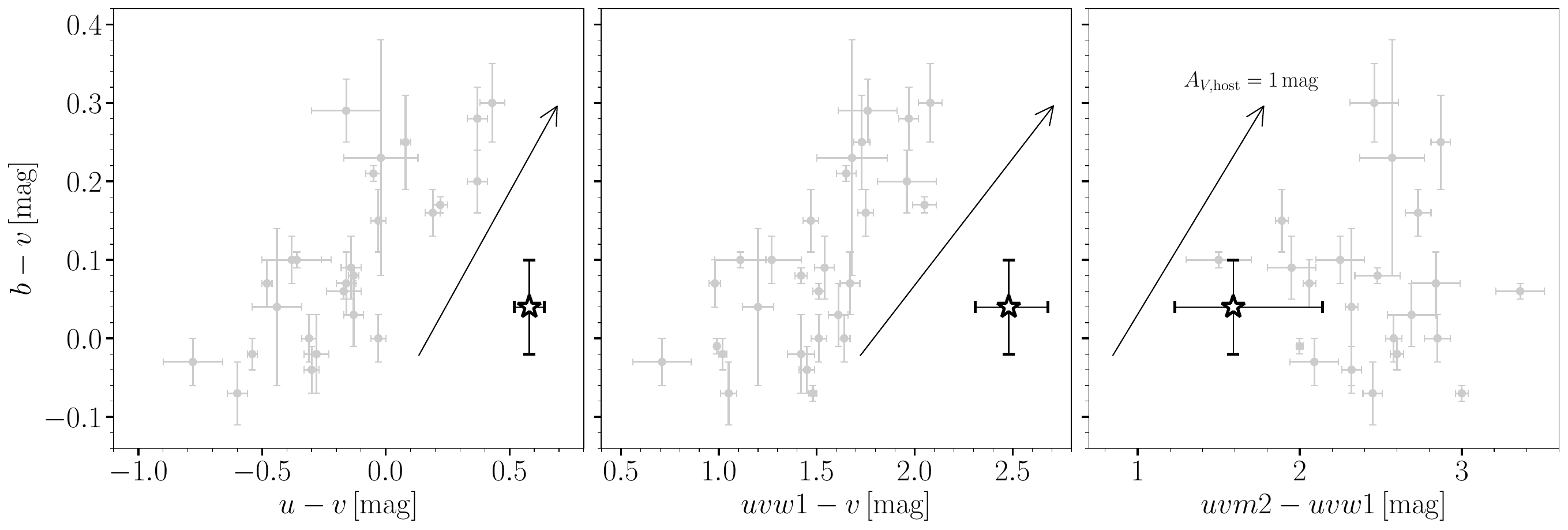}
    \caption{The color-color diagrams using UVOT photometry show that \sn\ (black star) has very unusual UV$-$optical colors at maximum luminosity compared to normal SNe\,Ia (gray dots; a sample of 29 normal SNe\,Ia from \citealp{Brown_2018}). The arrows mark how \sn\ would move in each color-color space if there were host reddening with the visual extinction of $A_{V,\mathrm{host}}=1$ (assuming $R_V=3.1$).
    }
    \label{fig:UVOT}
\end{figure*}

%UVOT color
While \sn\ shows a blue $g_\mathrm{ZTF}-r_\mathrm{ZTF}$ color near maximum brightness, its UV $-$ optical colors are unusually red. In Figure~\ref{fig:UVOT}, we show the locations of \sn\ in the UVOT color-color diagrams compared to those of 29 normal SNe\,Ia with UV observations around maximum from \citet{Brown_2018}. \sn\ stands out due to the red $u-v$ and $uvw1-v$ colors. After correcting for Galactic extinction, \sn\ shows {$u-v=0.58^{+0.06}_{-0.06}$\,mag and $uvw1 - v=2.48^{+0.23}_{-0.19}$\,mag}. As a comparison, none of the objects in the normal SN\,Ia sample has $u-v>0.5$\,mag or $uvw1 - v>2.1$\,mag. This cannot be a result of the unknown host reddening, since the amount of host extinction needed to account for the red near-UV colors of \sn\ would require the intrinsic $b-v$ color of the SN to be unphysically blue (shifted along the opposite direction of the arrows in Figure~\ref{fig:UVOT}). Interestingly, \sn\ exhibits a moderately blue mid-UV color ({$uvm2-uvw1=1.59^{+0.55}_{-0.36}$\,mag}), while most of the normal SNe\,Ia in the sample from \citet{Brown_2018} show $uvm2-uvw1\gtrsim2$\,mag. This might indicate that, for some reason, the flux in the near-UV ($\sim$2500--4000\,\r{A}) of \sn\ is suppressed near maximum brightness.

%conclusion
To conclude, despite a similar luminosity and color to 99aa-like/91T-like events at maximum brightness, the rapid photometric rise and decline and the unusual color evolution in \sn\ both indicate that it exhibits some peculiarities relative to normal and 99aa-like/91T-like SNe\,Ia.

\begin{deluxetable*}{lcccccc} \label{tab:basics}
\tabletypesize{\scriptsize}
\tablewidth{0pt}
\tablecaption{Basic photometric properties of \sn.}
\tablehead{
\multicolumn{7}{c}{Rise (flux $\le$$40\%$ of peak luminosity)}
}
\startdata
\multicolumn{3}{c}{Variable $\alpha$: prior -- $N(2,1)$} & & \multicolumn{3}{c}{Fixed $\alpha$: $\alpha=2$}\\
\cline{1-3} \cline{5-7}
$t_\mathrm{fl}$ (MJD) & $\alpha_{\mathrm{ZTF}, r}$ & $\alpha_{\mathrm{ATLAS}, o}$ &  & \multicolumn{3}{c}{$t_{\mathrm{fl}, \alpha=2}$ (MJD)}\\
$59703.16^{+0.70}_{-0.58}$ & $2.18^{+0.20}_{-0.24}$ & $2.37^{+0.48}_{-0.20}$ & & \multicolumn{3}{c}{$59703.66^{+0.10}_{-0.11}$}\\
\hline
\multicolumn{7}{c}{Maximum luminosity}\\[+0.1cm]
\hline
Filters & $g_\mathrm{ZTF}$ & $r_\mathrm{ZTF}$ & $B$ & $V$ & $R$ & $I$\\
\hline
$t_\mathrm{max,poly}$ (MJD) & $59722.66\pm0.21$ & $59725.54\pm0.09$ & $59722.77\pm0.30$ & $59724.88\pm0.28$ & $59724.61\pm0.28$ & $59720.73\pm0.27$\\
$M_\mathrm{max,poly}$ (mag) & $-19.693\pm0.014$ & $-19.492\pm0.004$ & $-19.456\pm0.011$ & $-19.544\pm0.009$ & $-19.496\pm0.009$ & $-19.222\pm0.011$\\
%$t_\mathrm{max,SALT}$ (MJD) & $59724.16\pm0.04$ & $59725.26\pm0.03$ & $59723.36\pm0.04$\\
%$M_\mathrm{max,SALT}$ (mag) & $-19.663\pm0.007$ & $-19.409\pm0.002$ & $-19.780\pm0.009$\\
\enddata
\tablecomments{Parameters are defined in the text. The absolute magnitudes have been corrected for Galactic extinction. The uncertainty in the distance modulus (0.03\,mag) and the systematics in the polynomial models are not included.}
\end{deluxetable*}

\subsection{Optical Spectral Properties} \label{sec:analysis_spec}
\begin{figure*}
    \centering
    \includegraphics[width=\linewidth]{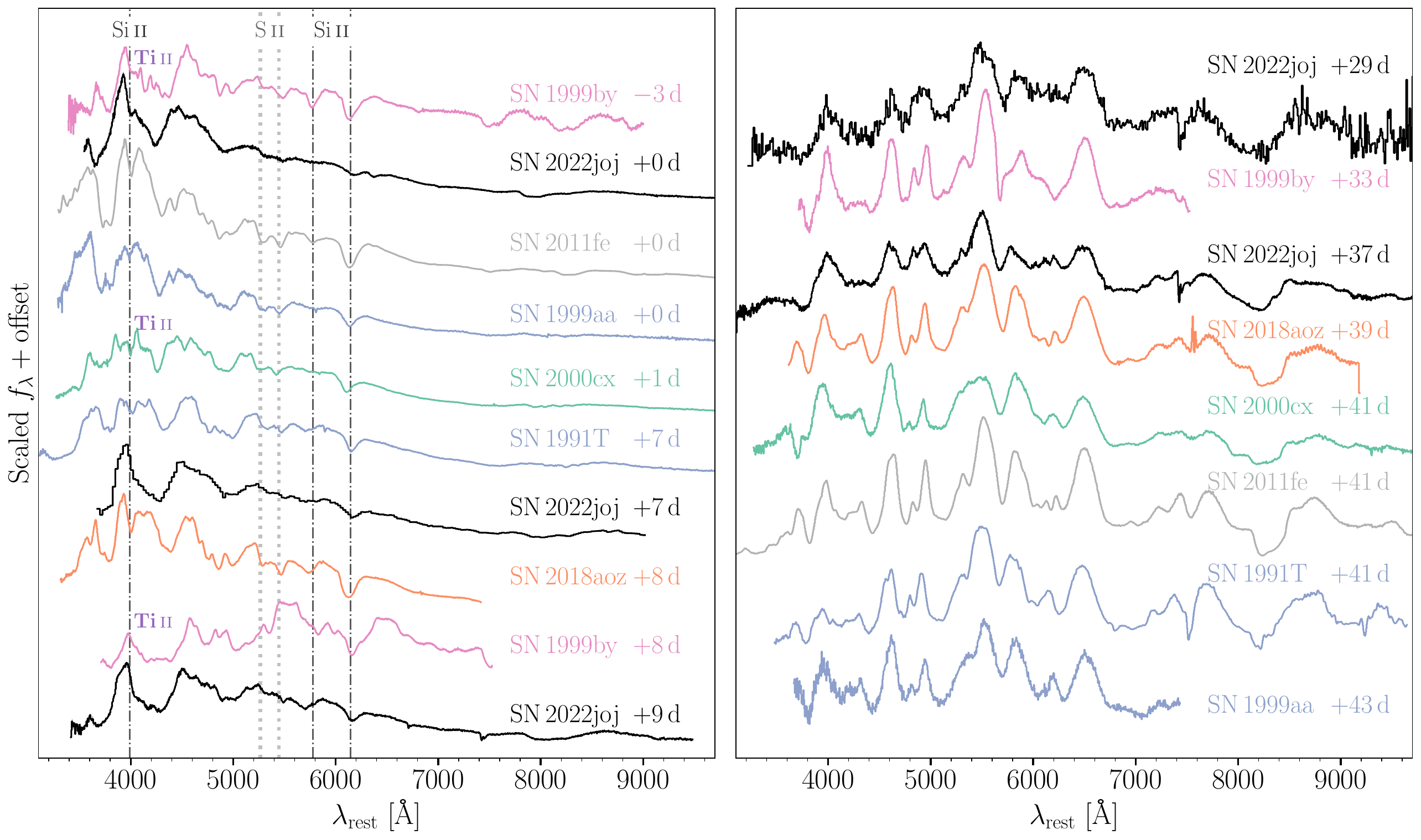}
    \caption{Optical spectra of \sn\ (black) and (i) a subluminous SN\,Ia, SN\,1999by (magenta), (ii) a normal SN\,Ia, SN\,2011fe (gray), (iii) two overluminous SNe\,Ia, SN\,1991T and SN\,1999aa (blue), (iv) the peculiar SN\,2000cx (green), and (v) a normal SN\,Ia with a red color at early times, SN\,2018aoz (orange), near maximum brightness (left panel) and about a month after maximum (right panel). The dash-dotted lines correspond to wavelengths of three \ion{Si}{2} lines (4128\,\r{A}, 5972\,\r{A}, and 6355\,\r{A}), while the dotted lines correspond to the wavelengths of the \ion{S}{2} ``W-trough'' (both assuming an expansion velocity of 10,000\,\kms). \ion{Ti}{2} has been identified from the spectra of SN\,1999by and SN\,2000cx at around $\sim$4200\,\r{A}, and the corresponding features are labeled. Spectra were downloaded from WiseREP  \citep{wiserep_2012}, with the following original data sources: SN\,2011fe -- \citet{Pereira_2013, Mazzali_2014}; SN\,1991T, SN\,1999aa, and SN\,2000cx -- \citet{Silverman_UCBIa_2012}; SN\,1999by -- \citet{Matheson_cfaIa_2008}; SN\,2018aoz -- \citet{Ni_18aoz_2023}.}
    \label{fig:spec_comp}
\end{figure*}
\begin{figure*}
    \centering
    \includegraphics[width=\linewidth]{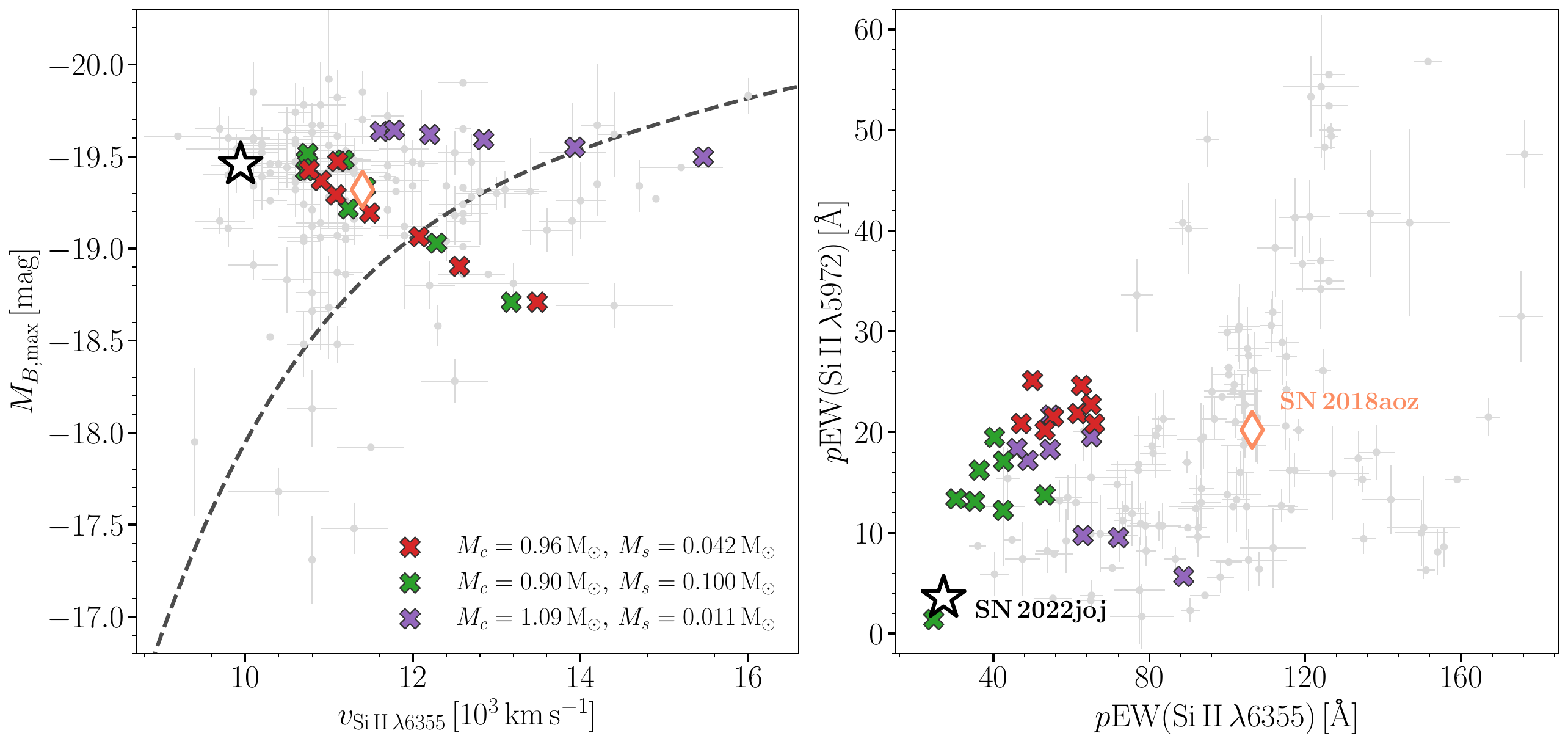}
    \caption{\sn\ (black star) is an SN\,Ia showing a normal brightness in $B$ and remarkably shallow \ion{Si}{2} features with a relatively low expansion velocity at maximum luminosity, compared to a sample of normal SNe\,Ia (gray dots) from \citet{Zheng_2018} and \citet{Burrow_2020}. \textit{Left:} the $B$-band absolute magnitude vs. the expansion velocity of \ion{Si}{2} $\lambda$6355 at maximum brightness. \textit{Right:} the pseudo-equivalent widths ($p$EWs) of the two \ion{Si}{2} lines, \ion{Si}{2} $\lambda\lambda$5972, 6355, at maximum brightness. The dashed black line corresponds to the theoretical $M_{B,\mathrm{max}}$-$v_\mathrm{Si\,\textsc{ii}}$ relation of 1D double-detonation models for thin helium shells across a spectrum of progenitor masses in \citet{polin_observational_2019}. The colored $\times$ symbols show 2D double-detonation models from \citet{Shen_2D_2021} with different C/O core mass ($M_c$) and helium shell mass ($M_s$), and viewing angles. For each model, multiple symbols are shown to summarize the effect of different viewing angles, from $\mu = -0.93$ to $\mu=+0.93$ ($\mu$ defined as the cosine of the polar angle relative to the point where the helium-shell detonation occurs). Parameters of the candidate double-detonation normal SN\,Ia, SN\,2018aoz \citep{Ni_18aoz_2023}, are also overplotted as an orange diamond.}
    \label{fig:phase_space}
\end{figure*}

In Figure~\ref{fig:spec_seq}, we show the optical spectral sequence of \sn. The $-12$\,days spectrum exhibits prominent absorption lines associated with \ion{Si}{2} $\lambda$6355 and \ion{Ca}{2} IRT (this spectrum was obtained and posted on TNS by \citealt{Newsome_2022TNSCR}). It also displays a strong suppression of flux blueward of $\sim$5000\,\r{A}, confirming the unusually red photometric colors at early times. Near maximum brightness, the Kast spectrum and the two SEDM spectra show a very blue continuum in the range $\sim$5000--8000\,\r{A} with shallow absorption features, indicating a high photometric temperature. \ion{Si}{2} $\lambda\lambda$5972, 6355 lines, the \ion{S}{2} W-trough, and \ion{Ca}{2} IRT are detectable but not prominent. The \ion{C}{2} $\lambda\lambda$6580, 7234 lines are prominent at maximum brightness ($+0$\,day), and quickly disappear afterward ($+3$\,days). A wide, asymmetric absorption feature appears at $\sim$4000--4500\,\r{A} (the 4200\,\r{A} features hereafter). There is a break on the blue edge of this feature which we associate with \ion{Si}{2} $\lambda$4128 that is often seen in other SNe\,Ia. The 4200\,\r{A} features become even wider and deeper in another SEDM spectrum at $+7$\,days and the ALFOSC spectrum at $+9$\,days. Weeks after the maximum, in the FLOYDS spectrum ($+29$\,days) and the LRIS spectrum ($+36$\,days), the bottom of the 4200\,\r{A} features becomes flat, reminiscent of the Ti-trough in the subluminous SN\,1991bg-like \citep[91bg-like;][]{Filippenko_91bg_1992,Leibundgut_91bg_1993} SNe. The nebular-phase spectra are dominated by [\ion{Fe}{2}] and [\ion{Fe}{3}] emission lines, but the [\ion{Fe}{2}] features (e.g, the complex around $\sim$7300\,\r{A}) are weaker than in other SNe\,Ia (Figure~\ref{fig:nebular_spec}), suggesting that the ejecta remain highly ionized about a year after the explosion. The nebular spectra are discussed in detail in Section~\ref{sec:disc_nebular}.

% 4200 A
In Figure~\ref{fig:spec_comp}, we compare maximum-light and transition-phase spectra of \sn\ to those seen in other SNe\,Ia. Around peak, the blue continuum and shallow absorption features in \sn\ are similar to those of overluminous objects, including SN\,1991T, SN\,1999aa, and SN\,2000cx. The asymmetric 4200\,\r{A} features are not seen in normal (SN\,2011fe) or overluminous (SN\,1991T and SN\,1999aa) SNe\,Ia, which all show another maximum at $\sim$4100\,\r{A} redward of the narrow \ion{Si}{2} $\lambda$4128 feature. In SN\,2000cx, a similar (but narrower) absorption feature is interpreted as high-velocity \ion{Ti}{2} \citep{Branch_00cx_2004}. The 4200\,\r{A} features are actually much more similar to the well-known ``Ti-trough'' that is ubiquitous in subluminous 91bg-like objects, e.g., SN\,1999by \citep{Arbour_1999}. Prior to the peak, SN\,1999by also shows this asymmetric absorption at about the same wavelength, which becomes more prominent with a nearly flat-bottom trough about a week after maximum. This absorption is caused by a blend of multiple species dominated by \ion{Ti}{2} \citep{Filippenko_91bg_1992,Mazzali_1997}. It remains prominent in the spectrum up to one month after maximum. Similarly, we find similar features in the spectra of \sn\ at $+29$\,days and $+36$\,days. Other normal/overluminous SNe\,Ia, unlike \sn, all exhibit a dip around $\sim$4500\,\r{A}. Aside from the 4200\,\r{A} features, \sn\ is otherwise entirely dissimilar from 91bg-like objects, which are $\gtrsim$2\,mag fainter at peak and exhibit much stronger \ion{Si}{2}, \ion{Ca}{2} and \ion{O}{1} absorption from a cooler line-forming region \citep{Filippenko_91bg_1992}. The Ti-trough in 91bg-like SNe is interpreted as the result of a low photospheric temperature \citep{Mazzali_1997}.

% shallow silicon
\sn\ also shows remarkably shallow \ion{Si}{2} absorption at maximum brightness. Following the techniques elaborated in \citet{Liu_20jgb_2023} \citep[see also][]{Childress_2013,Childress_2014,Maguire_2014}, we fit the \ion{Si}{2} and \ion{Ca}{2} IRT features with multiple Gaussian profiles. We find that modeling the \ion{Ca}{2} IRT absorption requires two distinct velocity components --- the photospheric-velocity features (PVFs) and the high-velocity features (HVFs). In Table~\ref{tab:vel_EW} we list the estimates of the expansion velocities and the pseudo-equivalent widths ($p$EWs) of the major absorption lines from $-12$\,days to $+9$\,days. In Figure~\ref{fig:phase_space} we show the peak absolute magnitude in $B$ band ($M_{B,\mathrm{max}}$) vs.\ the velocity and $p$EW of \ion{Si}{2} for \sn\ and a sample of normal SNe\,Ia from \citet{Zheng_2018} and \citet{Burrow_2020}. Figure~\ref{fig:phase_space} highlights that \sn\ has a normal $M_{B,\mathrm{max}}$ with a relatively low \ion{Si}{2} $\lambda$6355 expansion velocity (hereafter $v_\mathrm{Si\,\textsc{ii}}$). The \ion{Si}{2} $\lambda$5972 and \ion{Si}{2} $\lambda$6355 $p$EWs in \sn\ are lower than that for most normal SNe\,Ia. In fact, \sn\ sits at the extreme edge of the shallow-silicon group proposed in \citet{Branch_2006}, which mainly consists of overluminous 91T-like/99aa-like objects. This is consistent with the high luminosity and the blue color of \sn\ at maximum light, since a high photometric temperature results in higher ionization, reducing the abundance of singly ionized atoms (e.g., \ion{Si}{2}). Interestingly, the $p$EW of \ion{Si}{2} $\lambda$6355 near peak is significantly lower than that in the first spectrum. In typical 91T-like/99aa-like objects, the \ion{Si}{2} features are weak or undetectable at early times because the ejecta are even hotter, and only start to emerge around maximum light \citep{Filippenko_91T_1992}. In the early spectrum of \sn, in contrast, stronger absorption features from singly ionized Si and Ca indicate a cooler line-forming region at early times compared to that at maximum brightness.

% carbon
Prominent \ion{C}{2} features are also detected in the overluminous SN\,2003fg-like \citep[03fg-like;][]{Howell_2006} SNe\footnote{This subclass is also referred to as ``super-\Mch'' SNe or SN\,2009dc-like \citep[09dc-like;][]{Taubenberger_2011} SNe.} at their maximum brightness. However, 03fg-like objects show blue colors in the early light curves \citep{Taubenberger_2019} and appear even brighter in near-UV compared to normal SNe\,Ia at peak \citep{Brown_2014}, dissimilar from that of \sn. Many 03fg-like objects also show strong oxygen features both in their maximum-light and late-time spectra \citep{Taubenberger_2019}, while in \sn, we do not find evidence for oxygen. Consequently, the explosion mechanism as well as the origin of carbon in \sn\ is likely very different from that of 03fg-like objects.

In conclusion, the spectral evolution of \sn\ shows some similarities to 91T-like/99aa-like objects, as well as peculiarities. A reasonable model to explain \sn\ needs to reproduce (i) a strong suppression in flux blueward of $\sim$5000\,\r{A} at early times followed by a rapid evolution to blue colors; (ii) the seemingly contradictory observables at peak, namely the 4200\,\r{A} features similar to the Ti-trough in 91bg-like objects and the blue continuum/shallow \ion{Si}{2} feature at maximum, which separately indicate low and high photometric temperatures, respectively; and (iii) prominent \ion{C}{2} features at maximum brightness. 

\begin{deluxetable*}{lcccccccc} \label{tab:vel_EW}
\tabletypesize{\scriptsize}
\tablewidth{0pt}
\tablecaption{Fits to the expansion velocities and pEWs of \ion{Si}{2} $\lambda\lambda$5972, 6355 and the \ion{Ca}{2} IRT of \sn.}
\tablecolumns{9}
\tablehead{
\colhead{} &
\multicolumn{2}{c}{\ion{Si}{2} $\lambda$5972} &
\multicolumn{2}{c}{\ion{Si}{2} $\lambda$6355} &
\multicolumn{2}{c}{\ion{Ca}{2} IRT, PVFs} &
\multicolumn{2}{c}{\ion{Ca}{2} IRT, HVFs}\\
\cline{2-9}
\colhead{Phase} &
\colhead{$v$} & \colhead{$p$EW} &
\colhead{$v$} & \colhead{$p$EW} &
\colhead{$v$} & \colhead{$p$EW} &
\colhead{$v$} & \colhead{$p$EW}\\
\colhead{(day)} &
\colhead{($10^3$\,\kms)} & \colhead{(\AA)} &
\colhead{($10^3$\,\kms)} & \colhead{(\AA)} &
\colhead{($10^3$\,\kms)} & \colhead{(\AA)} &
\colhead{($10^3$\,\kms)} & \colhead{(\AA)}
}
\startdata
$-12.1$ & $\cdots$ & $\cdots$ & $-15.66\pm0.13$ & $47.5\pm2.5$ & $-14.85\pm0.83$ & $190\pm34$ & $-25.98\pm0.56$ & $278\pm40$ \\
$-0.3$ & $-10.10\pm0.88$ & $3.5\pm1.9$ & $-9.95\pm0.07$ & $27.2\pm0.8$ & $-12.45\pm0.17$ & $58\pm2$ & $-23.35\pm0.06$ & $117\pm2$ \\
$+2.5$ & $-8.77\pm0.63$ & $2.9\pm1.5$ & $-10.28\pm0.13$ & $27.8\pm1.3$ & $-12.03\pm0.73$ & $58\pm11$ & $-22.51\pm0.33$ & $109\pm11$ \\
$+2.6$ & $-8.35\pm0.62$ & $4.4\pm2.6$ & $-9.87\pm0.28$ & $25.4\pm2.7$ & $-11.77\pm0.93$ & $85\pm17$ & $-22.17\pm0.47$ & $105\pm15$\\
$+7.3$ & $\cdots$ & $\cdots$ & $-10.52\pm0.17$ & $40.9\pm2.7$ & $-10.10\pm0.82$ & $79\pm18$ & $-20.88\pm0.57$ & $172\pm24$\\
$+9.0$ & $\cdots$ & $\cdots$ & $-10.37\pm0.04$ & $48.5\pm0.6$ & $-11.94\pm0.23$ & $144\pm 6$ & $-21.13\pm0.15$ & $127\pm6$
\enddata
%\tablecomments{}
\end{deluxetable*}

\subsection{Host-Galaxy Properties}
We model the observed spectral energy distribution (SED; photometry in HSC-SSP $grizy$ and LS $W_1$ filters) with the software package \texttt{Prospector}\footnote{\texttt{Prospector} uses the \texttt{Flexible Stellar Population Synthesis} (\texttt{FSPS}) code \citep{Conroy_2009} to generate the underlying physical model and \texttt{python-fsps} \citep{ForemanMackey_FSPS_2014} to interface with \texttt{FSPS} in \texttt{python}. The \texttt{FSPS} code also accounts for the contribution from the diffuse gas based on the \texttt{Cloudy} models from \citet{Byler2017a}. We use the dynamic nested sampling package \texttt{dynesty} \citep{Speagle_dynesty_2020} to sample the posterior probability.} version 1.1 \citep{Johnson_prospector_2021}. The LS $grz$ photometry, which is consistent with the HSC-SSP results but has a lower S/N, is excluded from this modeling. We assume a Chabrier initial-mass function \citep[IMF;][]{Chabrier2003a} and approximate the star-formation history (SFH) by a linearly increasing SFH at early times followed by an exponential decline at late times (functional form $t \times \exp\left(-t/\tau\right)$, where $t$ is the age of the SFH episode and $\tau$ is the $e$-folding timescale). The model is attenuated with the \citet{Calzetti2000a} model. The priors of the model parameters are set identically to those used by \citet{Schulze2021a}. 

The SED is adequately described by a galaxy template with a mass of $\log(M_*/\mathrm{M_\odot}) = 7.13^{+0.15}_{-0.28}$, suggesting that the host is a dwarf galaxy. The modeled star-formation rate (SFR) is consistent with 0, but we note that measuring low SFRs with SED fitting is not robust and subject to systematics \citep{Conroy_2013}. In addition, the H$\alpha$ emission detected in the late-time spectrum of the SN (Section~\ref{sec:host}) indicates at least some level of star formation in the host.

\section{Discussion} \label{sec:discussion}
\subsection{\sn\ Compared to Model Explosions} \label{sec:model}

\begin{figure*}
    \centering
    \includegraphics[width=\linewidth]{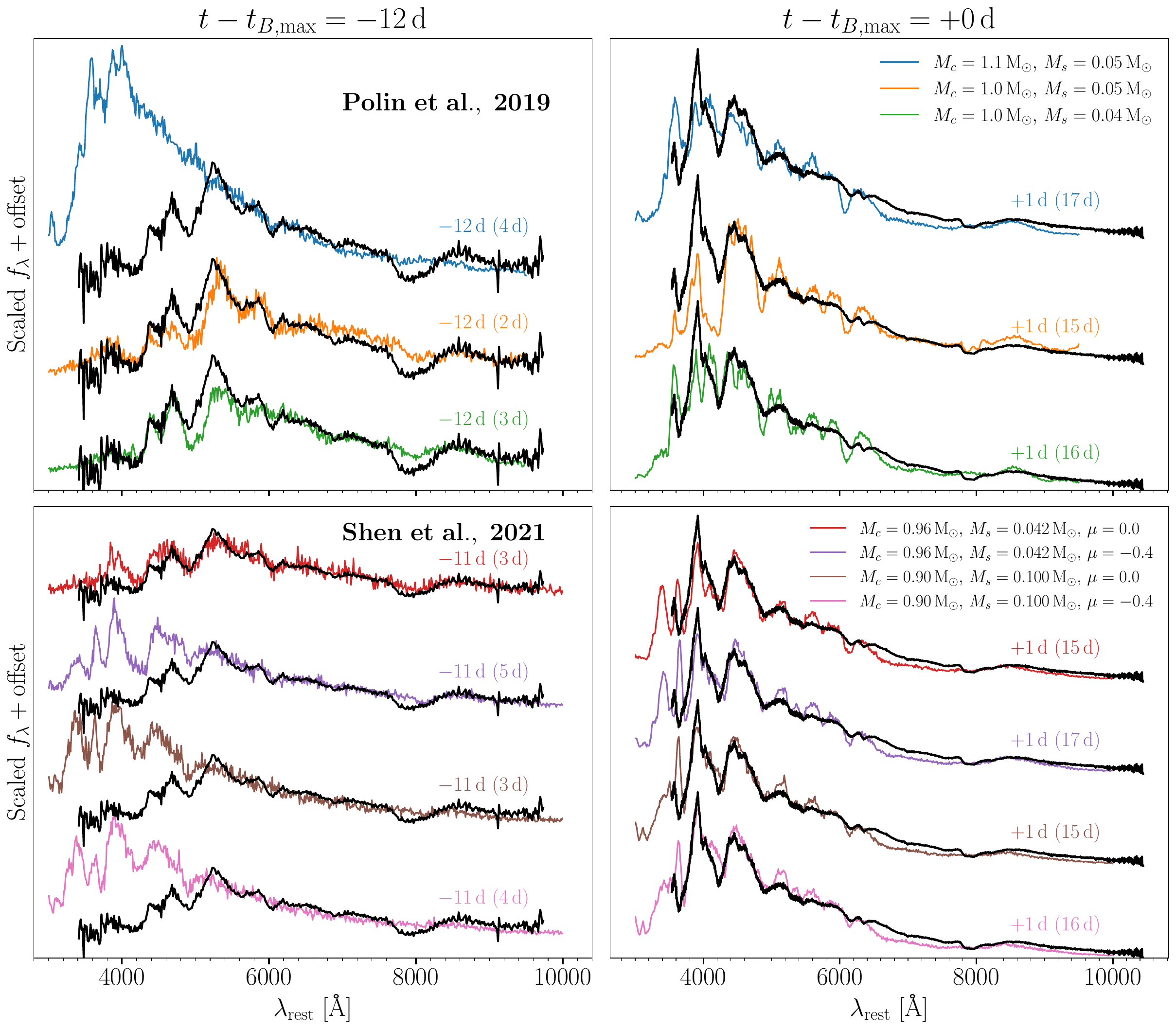}
    \caption{Comparisons of an early spectrum ($-12$\,days) and a maximum-light spectrum ($+0$\,days) of \sn\ (black) with two sets of double-detonation models. For each synthetic spectrum, the phase relative to $t_{B,\mathrm{max}}$  is listed (the time since explosion is shown in parentheses). \textit{Top:} comparison with 1D models from \citet{polin_observational_2019} with different C/O core ($M_c$) and He shell ($M_s$) masses. \textit{Bottom:} comparison with 2D models from \citet{Shen_2D_2021} with different masses and viewing angles $\mu$. The observed spectra have been corrected for Galactic extinction.
    }
    \label{fig:model_spec}
\end{figure*}

There are several physical mechanisms that can produce blue colors during the early evolution of SNe\,Ia, including heating of the SN ejecta following the decay of radioactive $^{56}$Ni, interaction of the SN ejecta with a nondegenerate companion \citep[e.g.,][]{Kasen_2010}, collisions between the SN ejecta and circumstellar material \citep[e.g.,][]{Piro_2016}, strong mixing that surfaces $^{56}$Ni to the outermost layers of the ejecta \citep[e.g.,][]{Piro_2013,Magee_2020}, and/or the production of radioactive isotopes in the detonation of a helium shell on the surface of the exploding WD \citep[e.g.,][]{Noebauer_2017,polin_observational_2019}.

In contrast, there are few proposed scenarios that can produce red colors up to a week after explosion, as in the case of \sn. If the newly synthesized $^{56}$Ni is strongly confined to the innermost SN ejecta, then an SN may remain red for several days after explosion as the heating diffuses out toward the photosphere \citep{Piro_2016}. Even the most confined $^{56}$Ni configuration considered in \citet{Piro_2016} converges to blue colors, similar to explosions with more extended $^{56}$Ni distributions, within $\sim$6\,days after explosion. \sn\ is observed to have very red colors $\sim$7\,days after $t_\mathrm{fl}$ (meaning more than 7\,days after explosion, since SNe\,Ia have a ``dark phase'' before photons diffuse out of the ejecta; \citealt{Piro_2013}). \citet{Dessart_2014} considered more realistic delayed-detonation scenarios. Some of their 1D unmixed delayed-detonation models still show a red $B-R$ color ($B-R\gtrsim0.5$\,mag) 7\,days after the explosion (see the DDC20, DDC22, and DDC25 models in their Figure~1), comparable to that of \sn. However, these models never appear as blue as \sn\ at peak ($B-R\simeq0.0$\,mag), and the $^{56}$Ni yields are relatively low ($M_\mathrm{Ni56}\lesssim0.3\,\mathrm{M_\odot}$), so they would result in subluminous events. In addition, we do not know any multidimensional explosion models that fail to produce any $^{56}$Ni mixing within the ejecta, and therefore disfavor this scenario.

Alternatively, in the double-detonation scenario, a layer of IGEs in the ashes of the helium shell can produce significant opacity in the outer layers of the bulk ejecta, producing a red color \citep{polin_observational_2019}. This scenario has been proposed for a few normal-luminosity SNe with red colors at early times, including SN\,2016jhr \citep{jiang_16jhr_2017} and SN\,2018aoz \citep{Ni_2022}. In Figure~\ref{fig:model_spec} we compare the spectra of \sn\ at $-12$\,days and $+0$\,day with 1D double-detonation models from \citet{polin_observational_2019} and 2D double-detonation models from \citet{Shen_2D_2021}. To create synthetic spectra, both models use \texttt{Sedona} \citep{Kasen_Sedona_2006}, a multi-dimensional radiative transfer (RT) simulator that assumes local thermodynamical equilibrium (LTE).

In the 1D models, the most important parameters are the mass of the C/O core ($M_c$) and the mass of the helium shell ($M_s$). The maximum luminosity depends on the amount of $^{56}$Ni synthesized in the explosion, which is predominantly determined by the total progenitor mass \citep[$M_c+M_s$;][]{polin_observational_2019}. We find that the maximum brightness in $B$ band ($M_{B,\mathrm{max}}=-19.46$\,mag) is reproduced by the 1D models with relatively massive progenitors ($\sim$1.1\,$\mathrm{M_\odot}$). However, models with such massive progenitors tend to produce blue, featureless spectra at early times (e.g., the $M_c=1.1\,\mathrm{M_\odot}$, $M_s=0.05\,\mathrm{M_\odot}$ model in Figure~\ref{fig:model_spec}), inconsistent with the observations. Less-massive models provide a better match to the line-blanketing seen in the early spectra, but fail to reproduce the maximum brightness as well as the 4200\,\r{A} features in the observed spectra. The 1D models overestimate the $p$EW and the expansion velocity of the \ion{Si}{2} $\lambda$6355 line at peak. As a reference, we overplot the theoretical $M_{B,\mathrm{max}}$-$v_\mathrm{Si\,\textsc{ii}}$ relation of 1D double-detonation models for thin helium shells ($M_s \simeq 0.01\,M_\odot$) across a spectrum of progenitor masses in \citet{polin_observational_2019} as the dashed black curve in the left panel of Figure~\ref{fig:phase_space}, which is not in agreement with the properties of \sn. Besides, none of the models produce detectable \ion{C}{2} features in the spectra at peak.

While the 1D models do not fully reproduce the observed properties of \sn, some of these tensions can be resolved when considering viewing-angle effects in multi-dimensional models. In Figure~\ref{fig:phase_space} we also show the properties of three 2D double-detonation models from \citet{Shen_2D_2021} with a variety of $M_c$ and $M_s$. For consistency in comparing the synthetic brightness with the observations of \sn, of which the $K$-corrections are unknown, the synthetic fluxes in the $B$ filter of these models are evaluated after shifting the synthetic spectra to $z=0.02736$, the redshift of \sn. We obtain the \ion{Si}{2} line properties using the same fitting techniques as in Section~\ref{sec:analysis_spec}.\footnote{The $v_\mathrm{Si\,\textsc{ii}}$ is systematically higher than the values displayed in Figure~20 from \citet{Shen_2D_2021}. In \citet{Shen_2D_2021}, the $v_\mathrm{Si\,\textsc{ii}}$ is determined using the minimum of the \ion{Si}{2} $\lambda$6355 absorption without subtracting the continuum, such that the estimated minimum is systematically redshifted with respect to the actual line center, whereas we fit the absorption features with Gaussian profiles on top of a linear continuum.}  
We again find that more-massive progenitors generally lead to higher-luminosity SNe, but different viewing angles produce significantly different spectral properties as a result of the asymmetry of the ejecta --- materials closer to the point of helium ignition are less dense and expand faster \citep{Shen_2D_2021}. In the plot, the cosine value of the polar angle relative to the point of helium ignition, $\mu$, ranges from $+0.93$ (near the helium ignition point) to $-0.93$ (opposite to the helium ignition point). When the SN is observed along a line of sight closer to the detonation point in the shell (greater $\mu$), it will appear fainter at maximum brightness and show a higher line velocity in \ion{Si}{2} $\lambda$6355. For a relatively high progenitor mass ($\gtrsim$$1.1\,\mathrm{M_\odot}$), a high $v_\mathrm{Si\,\textsc{ii}}$ ($\gtrsim$$13,500$\,\kms) is predicted in 1D models. However, all of the 2D models with $\mu\lesssim0$ show a lower $v_\mathrm{Si\,\textsc{ii}}$ ($\lesssim$$12,000$\,\kms), much closer to \sn\ in the $M_{B,\mathrm{max}}$-$v_\mathrm{Si\,\textsc{ii}}$ phase space. Models with $\mu>0$ are more consistent with 1D model predictions. It is suggested by \citet{polin_observational_2019} that high-velocity SNe that follow the dashed line in Figure~\ref{fig:phase_space} result from sub-\Mch\ double detonations, while SNe in the clump centered at $M_{B,\mathrm{max}}\simeq-19.5$\,mag and $v_\mathrm{Si\,\textsc{ii}}\simeq11,000$\,\kms\ are likely near-\Mch\ explosions. Based on the 2D models, however, we should expect a similar number of high-velocity and normal-velocity double-detonation SNe\,Ia. A substantial fraction of the objects within the clump on the $M_{B,\mathrm{max}}$-$v_\mathrm{Si\,\textsc{ii}}$ diagram may be sub-\Mch\ double-detonation events viewed from certain orientations \citep{Shen_2D_2021}. SN\,2018aoz is a double-detonation candidate that, like \sn, exhibits early red colors before evolving to normal luminosity and blue colors \citep{Ni_2022}. Interestingly, SN\,2018aoz also resides in the high-luminosity, low-velocity clump in the $M_{B,\mathrm{max}}$-$v_\mathrm{Si\,\textsc{ii}}$ space \citep{Ni_18aoz_2023}, and thus, it too may be an example of a double-detonation SN\,Ia viewed from the hemisphere opposite to the helium ignition point.

In the bottom panels of Figure~\ref{fig:model_spec} we show two 2D double-detonation models (each with two viewing angles) with a total progenitor mass of $\sim$$1\,\mathrm{M_\odot}$ from \citet{Shen_2D_2021}. These models qualitatively match the observed spectra at maximum light. In the $M_c=0.96\,\mathrm{M_\odot}$, $M_s=0.042\,\mathrm{M_\odot}$ model, we find that when $\mu=0$ (viewed from the equator), it predicts a reasonable level of line blanketing in the blue side of the spectrum at early times (lower-left panel of Figure~\ref{fig:model_spec}). Near maximum brightness, the model also reproduces the overall shape of the observed spectrum, though the strength of nearly all the absorption lines (4200\,\r{A} features, \ion{S}{2}, and \ion{Si}{2}) is overestimated, and $v_\mathrm{Si\,\textsc{ii}}$ is also overestimated ($\sim$12,000\,\kms). When viewed from the hemisphere opposite to the helium ignition point (e.g., $\mu=-0.4$), the model yields an asymmetric profile of 4200\,\r{A} features that matches the observations better. The \ion{Si}{2} features are also predicted to be shallower, though still not as shallow as that in the observations. Nonetheless, the spectra at early times are expected to be much bluer than the observations. The $M_c=0.90\,\mathrm{M_\odot}$, $M_s=0.100\,\mathrm{M_\odot}$ models, especially when $\mu=-0.4$, produce even shallower and slower-expanding \ion{Si}{2} features at maximum brightness. This is in agreement with the trend observed in the right panel of Figure~\ref{fig:phase_space}: models with a thicker helium shell tend to exhibit shallower \ion{Si}{2} features. However, the level of line blanketing blueward of $\sim$5000\,\r{A} is also underestimated at early times. In addition, none of the models produce significant \ion{C}{2} features, though we will show in Section~\ref{sec:disc_C_II} that this discrepancy does not necessarily invalidate the double-detonation interpretation.

To investigate the origin of the 4200\,\r{A} features at maximum luminosity, we run additional 1D \texttt{Sedona} RT simulations for the $M_c=0.96\,\mathrm{M_\odot}$ and $M_s=0.042\,\mathrm{M_\odot}$ from \citet[adopted in the calculations of \citealp{Shen_2D_2021}]{Boos_2021}. We adopt the density and chemical profile of the slice between the viewing angles $\mu=-0.467$ and $\mu=-0.333$ in the 2D ejecta at $t_{B,\mathrm{max}}$ estimated in the original 2D RT simulations (16.25\,days after the explosion) as the input of the 1D model. The synthetic spectrum is generally consistent with the 2D RT outcomes with a viewing angle $\mu=-0.4$. Then we run another 1D simulation with all the titanium isotopes in the slice removed. The resultant synthetic spectrum is still broadly consistent with the 1D and 2D results redward to $\sim$5000\,\r{A}, but shows a significant peak at $\sim$4100\,\r{A} resembling those in normal SNe\,Ia. This indicates that an explosion that yields an extended titanium distribution in the ejecta (such as the example double-detonation model shown here), a blend of Ti lines reshapes the spectrum around 4200\,\r{A}, leading to the absence of the peak at $\sim$4100\,\r{A} and a deep, asymmetric absorption feature.

\begin{figure}
    \centering
    \includegraphics[width=\linewidth]{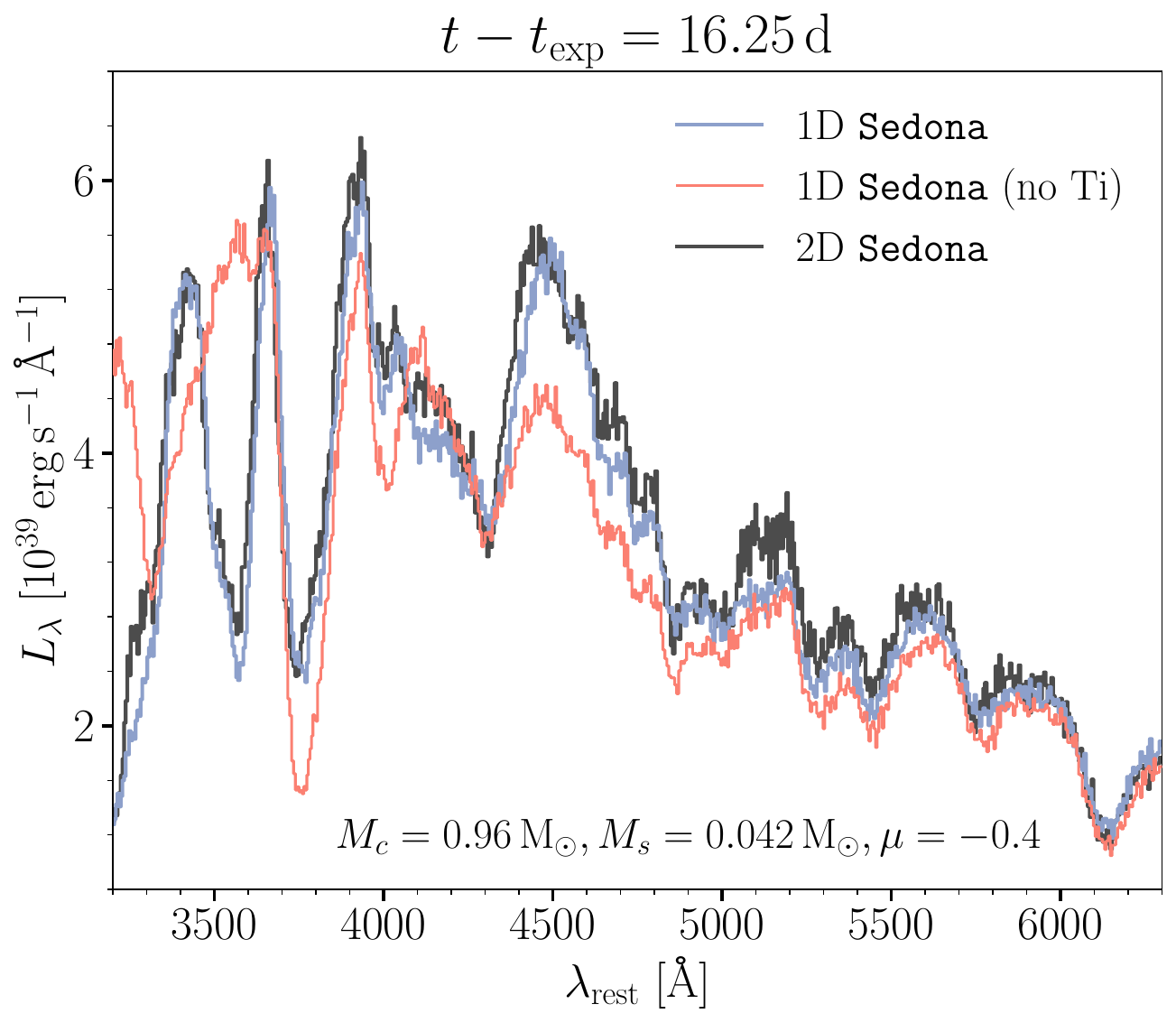}
    \caption{Comparison of the \texttt{Sedona} synthetic spectra at maximum luminosity indicates that Ti dominates the 4200\,\r{A} features in a double-detonation model. Two 1D \texttt{Sedona} models (blue + salmon pink) are run using the density and chemical profile of a slice between the viewing angles $\mu=-0.467$ and $\mu=-0.333$ of the $M_c=0.96\,\mathrm{M_\odot}$ and $M_s=0.042\,\mathrm{M_\odot}$ model from \citet{Boos_2021}. While the synthetic spectrum in the original 2D \texttt{Sedona} model can be well reproduced in a 1D run (blue), the model with all the titanium isotopes removed (salmon pink) cannot reproduce the remarkable 4200\,\r{A} features and exhibits a peak at $\sim$4100\,\r{A}.}
    \label{fig:Ti}
\end{figure}

The extremely red UV-optical colors near maximum luminosity are also broadly consistent with the double-detonation scenario, since the heavy elements (e.g., Ti, V, Cr, and Fe) in the outer ejecta could effectively absorb the UV photons with wavelengths around 3000\,\r{A}. 
However, no existing models could accurately model the UVOT light curves, especially in the $u$-band ($\sim$3100--3900\,\r{A} in the observed frame, or $\sim$3000--3800\,\r{A} in the rest frame of the host galaxy), where the flux is dominated by the re-emission of Ti, V, and Cr, and non-LTE effects could be important.

While none of the models presented here provide a strong match to \sn\ at every phase, we draw the broad conclusion that the spectroscopic properties of \sn\ are qualitatively consistent with a sub-\Mch\ WD ($\gtrsim$1.0\,$\mathrm{M_\odot}$) double detonation viewed from the hemisphere opposite to the ignition point. Observers from such a viewing angle would observe strong absorption features in the blue portion of the spectrum dominated by Ti as well as relatively shallow and slowly-expanding Si lines in the red portion. We emphasize that none of the models considered here was specifically developed and tuned to explain \sn. Customized models specifically tuned for \sn\ may reproduce all the observed features simultaneously, and we suggest more 2D double detonation simulations be performed. Furthermore, additional improvements can be made via an improved handling of the radiative transfer \citep[e.g., non-LTE effects; see][]{Shen_NLTE_2021}.

\subsection{The 7300\,\r{A} Region in Nebular-Phase Spectra}
\label{sec:disc_nebular}
\begin{figure*}
    \centering
    \includegraphics[width=\linewidth]{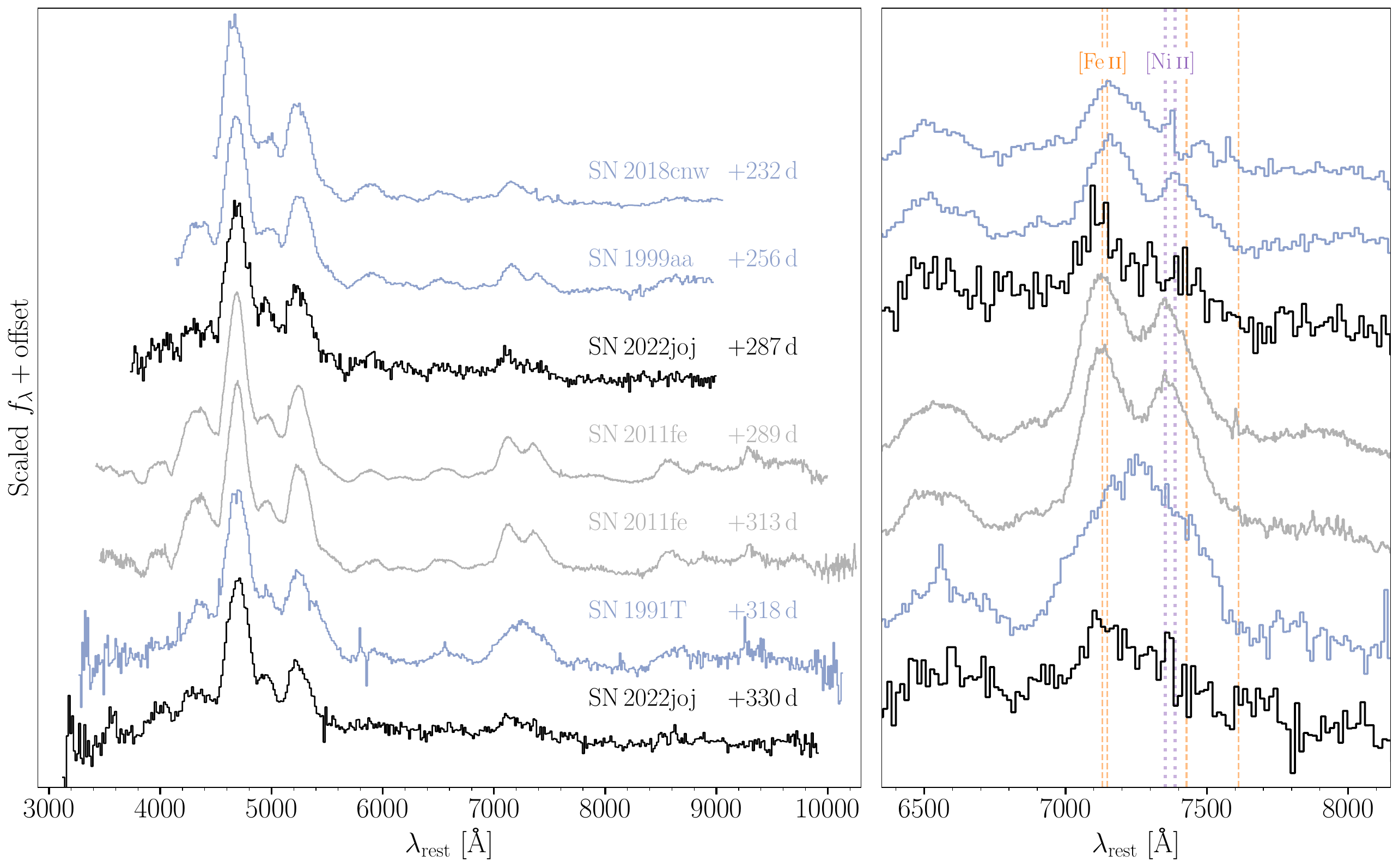}
    \caption{Nebular-phase spectra of \sn\ (black), three overluminous SNe\,Ia (blue), SN\,1991T, SN\,1999aa, and SN\,2018cnw, and a normal SN\,Ia (gray), SN\,2011fe. The right panel zooms in on the features around 7300\,\r{A}. The flux has been normalized to the [\ion{Fe}{3}] features around 4700\,\r{A}. The orange dashed lines correspond to wavelengths of four [\ion{Fe}{2}] lines (7155\,\r{A}, 7172\,\r{A}, 7453\,\r{A}, and 7638\,\r{A}), while the purple dotted lines correspond to the wavelengths of two [\ion{Ni}{2}] lines (7378\,\r{A}, 7412\,\r{A}), both blueshifted by 1000\,\kms. Spectra were downloaded from WiseREP \citep{wiserep_2012}, with the following original data sources: SN\,1991T and SN\,1999aa -- \citet{Silverman_UCBIa_2012}; SN\,2011fe -- \citet{Mazzali_2015}; SN\,2018cnw -- this work.}
    \label{fig:nebular_spec}
\end{figure*}
\begin{figure}
    \centering
    \includegraphics[width=\linewidth]{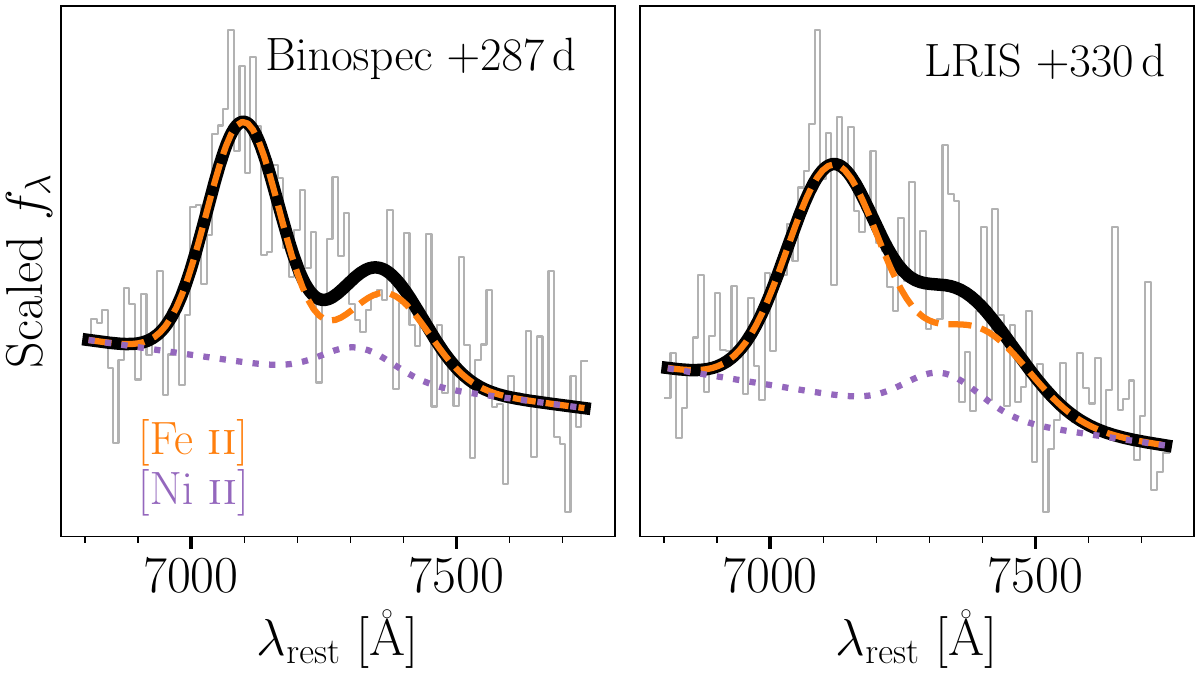}
    \caption{Fits to the 7300\,\r{A} region containing [\ion{Fe}{2}] and [\ion{Ni}{2}] features are consistent with a low \ion{Ni}{2} abundance. The observed spectra are shown in gray. The colored lines correspond to the models of [\ion{Fe}{2}] (orange) and [\ion{Ni}{2}] (purple) features. For the model parameters we adopt the mean values of their posterior distributions. The black solid lines are the overall models.}
    \label{fig:Fe_ni}
\end{figure}

In Figure~\ref{fig:nebular_spec} we compare the two nebular-phase spectra of \sn\ with the overluminous SNe\,Ia (SN\,1991T, SN\,1999aa, and SN\,2018cnw) and the normal-luminosity SN\,2011fe.

Compared to that of other SNe\,Ia, the nebular spectra of \sn\ show a relatively low flux ratio between the complex at $\sim$7300\,\r{A} (hereafter the 7300\,\r{A} features) dominated by [\ion{Fe}{2}] and [\ion{Ni}{2}] and the complex at $\sim$4700\,\r{A} dominated by [\ion{Fe}{3}]. This suggests high ionization in the ejecta \citep{Wilk_2020}. In addition to the smaller flux ratio, the profile of the 7300\,\r{A} features in \sn\ is also distinct from other SNe. Most of SNe\,Ia show a bimodal structure in their 7300\,\r{A} features \citep[e.g.,][]{Graham_2017,Maguire_2018}. The bluer peak is dominated by [\ion{Fe}{2}] $\lambda\lambda$7155, 7172, while [\ion{Ni}{2}] $\lambda\lambda$7378, 7412 usually have non-negligible contributions to the redder peak (see Figure~\ref{fig:nebular_spec}). In some peculiar SNe\,Ia (mostly subluminous ones), the detection of [\ion{Ca}{2}] $\lambda\lambda$7291, 7324 has also been reported \citep[e.g.][]{jacobson-galan_16hnk_2020,Siebert_19yvq_2020}. The bimodal morphology is prominent in the spectra of SN\,1999aa and SN\,2011fe. SN\,1999T is well-known for its broader emission lines in the nebular phase, so the composition of the 7300\,\r{A} features is ambiguous. In the spectra of \sn\ and SN\,2018cnw, however, the redder peak is absent and the 7300\,\r{A} features show an asymmetric single peak, which seems to indicate a low abundance of Ni in the ejecta.

To investigate the relative contributions of [\ion{Fe}{2}] and [\ion{Ni}{2}] to the 7300\,\r{A} features in \sn, we model this region with multiple Gaussian emission profiles using the same technique as in Section~\ref{sec:analysis_spec}. We include four [\ion{Fe}{2}] lines (7155, 7172, 7388, 7453\,\r{A}) and two [\ion{Ni}{2}] lines (7378, 7412\,\r{A}) in the fit. For each species, the relative flux ratios of lines are fixed, whose values are adopted from \citet{Jerkstrand_2015}. For [\ion{Fe}{2}], we set $L_{7155}:L_{7172}:L_{7388}:L_{7453} = 1:0.24:0.19:0.31$, and for [\ion{Ni}{2}], we set $L_{7378}:L_{7412} = 1:0.31$. These line ratios are calculated assuming LTE, but the departure from LTE should not be significant under the typical conditions in the ejecta \citep{Jerkstrand_2015}. We allow the amplitudes $A$ of these Gaussian profiles to be either positive or negative. The velocity dispersions $\sigma_v$ in different lines of each species are set to be the same. For both $A$ and $\log\sigma_v$, we adopt flat priors, and only allow $\sigma_v$ to vary between 1000 and 6000\,\kms. In addition, we adopt wide Gaussian priors for the radial velocities $v$ of [\ion{Fe}{2}] and [\ion{Ni}{2}], both centered at $-1000$\,\kms\ with a standard deviation of $2000$\,\kms. The fitted models are shown in Figure~\ref{fig:Fe_ni}, where colored curves correspond to the [\ion{Fe}{2}] and [\ion{Ni}{2}] emission adopting the mean values of the posterior distributions of model parameters sampled with MCMC. The fitted parameters are listed in Table~\ref{tab:Fe_Ni}. In both spectra, the flux of [\ion{Ni}{2}] is consistent with 0 (${L_{[\mathrm{Ni}\,\textsc{ii}]\,\lambda7378}}/{L_{[\mathrm{Fe}\,\textsc{ii}]\,\lambda7155}}=0.13\pm0.14$ at $+287$\,days and $0.18\pm0.16$ at $+330$\,days; the flux ratios of lines are estimated with the ratios of their $p$EWs, and the uncertainties are the robust standard deviations estimated with the median absolute deviation of the drawn sample). We also find that the [\ion{Ni}{2}] velocities cannot be constrained by the data, whose uncertainties estimated in both spectra are greater than the standard deviations of their Gaussian priors (2000\,\kms), again disfavoring a solid [\ion{Ni}{2}] detection. The 7300\,\r{A} features can be well fit with [\ion{Fe}{2}] emission only. In addition, we have tested fitting this complex with [\ion{Ca}{2}] in addition to [\ion{Fe}{2}] and [\ion{Ni}{2}], but find no evidence for [\ion{Ca}{2}].

The relative abundance of Ni and Fe, which probes the mass of the progenitor WD, can be estimated via the flux ratio of their emission lines. At $>$300\,days after explosion $^{56}$Fe is the dominant isotope of Fe following the decay of $^{56}$Ni through the chain $^{56}$Ni$\rightarrow^{56}$Co$\rightarrow^{56}$Fe. Consequently, the Fe abundance primarily depends on the yield of $^{56}$Ni. The Ni abundance, however, is sensitive to both the progenitor mass and the explosion scenario. The stable Ni isotopes ($^{58}$Ni, $^{60}$Ni, and $^{62}$Ni) are more neutron-rich compared to the $\alpha$-species $^{56}$Ni, and can only be formed in high-density regions with an enhanced electron-capture rate during the explosion \citep{Nomoto_1984,Khokhlov_1991}. Consequently, SNe\,Ia from sub-\Mch\ WDs, with central densities that are lower than near-\Mch\ WDs, are expected to show a lower abundance of stable Ni isotopes \citep{Iwamoto_1999,Seitenzahl_2013,Shen_DD_2018}. 

To estimate the relative abundance of Ni and Fe, we use the equation adopted in \citet{Jerkstrand_2015} and \citet{Maguire_2018},
\begin{equation}
    \frac{L_{7378}}{L_{7155}} = 4.9\frac{n_{\mathrm{Ni}\,\textsc{ii}}}{n_{\mathrm{Fe}\,\textsc{ii}}}\exp\left(\frac{0.28\,\mathrm
    {eV}}{k_BT}\right)\frac{dc_{\mathrm{Ni}\,\textsc{ii}}}{dc_{\mathrm{Fe}\,\textsc{ii}}},
\end{equation}
where $L_{7378}/L_{7155}$ is the flux ratio of the [\ion{Ni}{2}] $\lambda$7378 to [\ion{Fe}{2}] $\lambda$7155 lines, $n_{\mathrm{Ni}\,\textsc{ii}}/n_{\mathrm{Fe}\,\textsc{ii}}$ is the number density ratio of \ion{Ni}{2} and \ion{Fe}{2}, and ${dc_{\mathrm{Ni}\,\textsc{ii}}}/{dc_{\mathrm{Fe}\,\textsc{ii}}}$ is the ratio of the departure coefficients from LTE for these two ions. Since both \ion{Ni}{2} and \ion{Fe}{2} are singly ionized species with similar ionization potentials, we assume that $n_{\mathrm{Ni}\,\textsc{ii}}/n_{\mathrm{Fe}\,\textsc{ii}}$ is a good approximation of the total Ni/Fe ratio. As illustrated in \citet{Maguire_2018}, this assumption proves to be valid by modeling nebular-phase spectra at similar phases \citep{Fransson_2015,Shingles_2022}, with the relative deviation from the ionization balance $\lesssim$20\%. We handle the uncertainties due to the unknown temperature, ratio of departure coefficients, and ionization balance in a Monte Carlo way. We randomly generate $N=4000$ samples of the temperature (3000--8000\,K), the ratio of departure coefficients (1.2--2.4), and the ionization balance factor (0.8--1.2) assuming uncorrelated uniform distributions. These intervals are again adopted from \citet{Maguire_2018}. Combining these quantities with the samples of line-profile parameters drawn with the MCMC, we obtain $N$ estimates of Ni/Fe, which are effectively drawn from its posterior distribution. We find that Ni/Fe is consistent with 0 and we obtain a 3$\sigma$ upper limit of $\mathrm{Ni/Fe}<0.03$. Such a low Ni abundance is more consistent with the yields of sub-\Mch\ double-detonation scenarios \citep{Shen_DD_2018}, much lower than the expected outcomes of near-\Mch, delayed-detonation models \citep{Seitenzahl_2013} or pure deflagration models \citep{Iwamoto_1999}.

Alternatively, it is proposed in \citet{Blondin_2022} that for high-luminosity SNe\,Ia, the absence of [\ion{Ni}{2}] lines can be a result of high ionization of Ni in the inner ejecta, despite the fact that a significant amount of Ni exists. It is shown that the [\ion{Ni}{2}] $\lambda\lambda$7378, 7412 lines can be strongly suppressed even in a high-luminosity, near-\Mch\ delayed-detonation model, once the \ion{Ni}{2}/\ion{Ni}{3} ratio at the center of the ejecta is artificially reduced by a factor of 10. Nevertheless, it remains to be questioned whether a physical mechanism exists to boost the ionization in the inner ejecta, where the stable Ni dominates the radioactive $^{56}$Ni and $^{56}$Co, and the deposited energy per particle due to the radioactive decay is usually low. One possible scenario is inward mixing, which brings $^{56}$Co into the innermost ejecta such that the ionization would significantly increase. However, in this case calcium would inevitably be mixed inward as well, and the resultant \ion{Ca}{2} $\lambda\lambda$7291, 7324 lines would stand out and dominate the 7300\,\r{A} features \citep{Blondin_2022}. Other physical mechanisms are thus required to reduce the \ion{Ni}{2}/\ion{Ni}{3} ratio at the center of a near-\Mch\ explosion.

We also find that the [\ion{Fe}{2}] lines are significantly blueshifted ($v_\mathrm{[Fe\,\textsc{ii}]}=-(2.46\pm0.38)\times10^3\mathrm{\,km\,s^{-1}}$ at $+287$\,days and $-(1.56\pm0.60)\times10^3\mathrm{\,km\,s^{-1}}$ at $+330$\,days). This is consistent with other SNe\,Ia showing low $v_\mathrm{Si\,\textsc{ii}}$ at maximum brightness \citep{Maeda_2010,Maguire_2018,Li_2021}, and also in qualitative agreement with the asymmetric sub-\Mch\ double-detonation scenario. Specifically, along a line of sight opposite to the shell detonation point, observers would see intermediate-mass elements (IMEs) with low expansion velocities, including \ion{Si}{2}. In the meantime, the IGEs at the center of the ejecta would have a bulk velocity toward the observer (see Figure~1 and Figure~2 of \citealp{Bulla_2016}; see also \citealp{Fink_DD_2010}).

\begin{deluxetable*}{lcccccc} \label{tab:Fe_Ni}
\tabletypesize{\scriptsize}
\tablewidth{0pt}
\tablecaption{Fits to the late-time spectra around 7300\,\r{A} with [\ion{Fe}{2}] and [\ion{Ni}{2}] emission.}
\tablecolumns{7}
\tablehead{
\colhead{} &
\multicolumn{3}{c}{[\ion{Fe}{2}] $\lambda$7155} &
\multicolumn{3}{c}{[\ion{Ni}{2}] $\lambda$7378}\\
\cline{2-7}
\colhead{Phase} & 
\colhead{$v$} & \colhead{$\sigma_v$} & \colhead{$p$EW} &
\colhead{$v$} & \colhead{$\sigma_v$} &\colhead{$p$EW} \\
\colhead{(day)} &
\colhead{($10^3$\,\kms)} & \colhead{($10^3$\,\kms)} & \colhead{(\AA)} &
\colhead{($10^3$\,\kms)} & \colhead{($10^3$\,\kms)} & \colhead{(\AA)}
}
\startdata
$+286.8$ & $-2.46\pm0.38$ & $2.98\pm0.47$ & $-129\pm25$ & $-2.89\pm2.60$ & $3.02\pm1.45$ & $-14\pm19$ \\
$+329.8$ & $-1.56\pm0.60$ & $3.84\pm0.80$ & $-113\pm28$ & $-2.49\pm2.44$ & $3.11\pm1.41$ & $-13\pm22$\\
\enddata
%\tablecomments{}
\end{deluxetable*}

\subsection{Carbon Features at Maximum Luminosity}\label{sec:disc_C_II}
\begin{figure}
    \centering
    \includegraphics[width=\linewidth]{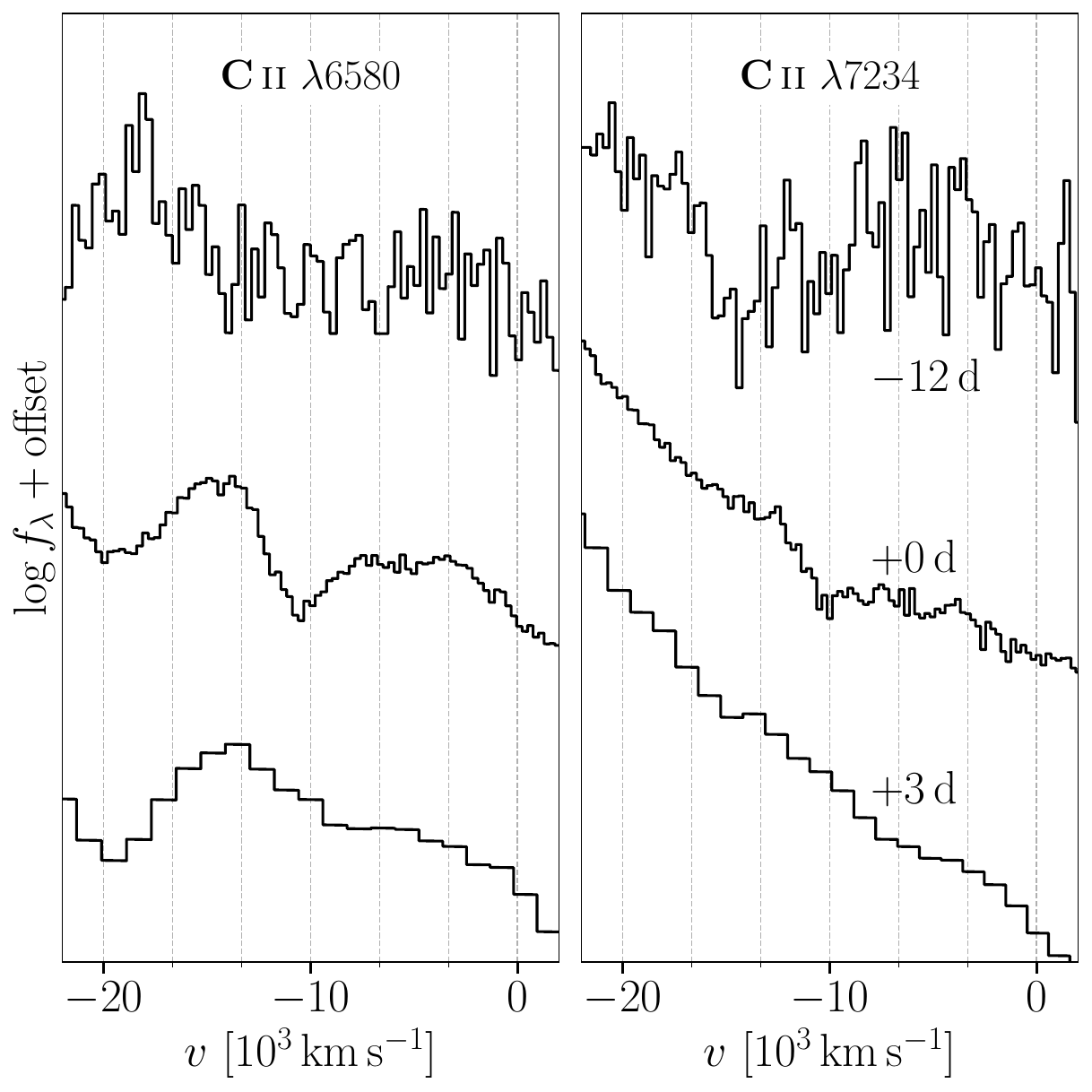}
    \caption{Prominent \ion{C}{2} $\lambda\lambda$6580, 7234 absorption lines at maximum luminosity. No evident \ion{C}{2} $\lambda$6580 lines are found in the earliest spectrum ($-12$\,days; though there is one absorption feature that might be associated with \ion{C}{2} $\lambda$7234 at $\sim$14,000\,\kms) or the post-maximum spectrum ($+3$\,days). All spectra are binned with a bin size of 5\,\AA, except for the low-resolution SEDM spectrum ($+3$\,days). The dashed lines correspond to wavelengths of the \ion{C}{2} lines assuming a range of velocities.}
    \label{fig:C_II}
\end{figure}
The spectrum at maximum luminosity exhibits strong \ion{C}{2} $\lambda\lambda$6580, 7234 lines at a velocity of $\sim$10,000\,\kms\  (Figure~\ref{fig:C_II}). In the $-12$\,days spectrum, we do not find strong evidence for the \ion{C}{2} $\lambda$6580 line, though there is an absorption feature that might be associated with the \ion{C}{2} $\lambda$7234 line at an expansion velocity of $\sim$$14,000$\,\kms (see the right panel in Figure~\ref{fig:C_II}). The feature shows a velocity dispersion that is a factor of $\sim$3 greater than the unambiguous \ion{C}{2} $\lambda$7234 line at maximum luminosity, and its nature remains vague. In this section we briefly discuss the possible origin of the carbon features.

In double detonations of sub-\Mch\ WDs, the burning of carbon is expected to be efficient, with only a small fraction of carbon left unburnt. In 1D double-detonation models from \citet{polin_observational_2019}, only a negligible ($\lesssim$$10^{-5}\,\mathrm{M_\odot}$) amount of carbon is unburnt in normal and overluminous SNe, which should not cause any noticeable spectroscopic features. Multidimensional simulations produce a greater amount ($\sim$$10^{-4}$--$10^{-3}$\,$\mathrm{M_\odot}$) of leftover carbon \citep{Fink_DD_2010,Boos_2021}, which is still roughly two orders of magnitude lower than the yields of IMEs. Nevertheless, most of the unburnt carbon will be concentrated on the surface of the C/O core, forming a sharp carbon-rich shell \citep[see Figures~5--8 in][]{Boos_2021} with a high expansion velocity ($>$10,000\,\kms). Both 1D and 2D double-detonation models have low carbon yields in the helium ashes, so the carbon features may be strengthened over time, as the photosphere moves inward to the carbon-rich regions. Nevertheless, with the current resolution in the RT simulations, the contribution of such a sharp shell is unlikely to be captured in the resultant synthetic spectra. 

Another possible origin of unburnt carbon in a double-detonation SN is the stripped material from a degenerate companion. In the Dynamically Driven Double-Degenerate Double-Detonation (D$^6$) model \citep{Shen_2018}, the primary C/O WD could detonate following the dynamical ignition of a helium shell from its WD companion (either a helium WD or a C/O WD with a substantial helium envelope) during the unstable mass transfer \citep{Guillochon_2010,Pakmor_2013}. In this case, a significant amount ($\sim$$10^{-3}\,\mathrm{M_\odot}$) of materials can be stripped from a C/O WD companion, following the explosion of a 1\,$\mathrm{M_\odot}$ primary WD (\citealp{Tanikawa_2018}; Boos et al. 2023, in prep.), although the amount of carbon would likely be significantly less if the companion still holds a substantial helium envelope \citep{Tanikawa_2019}. The velocity of the stripped carbon is expected to be low \citep[e.g., centered at $\sim$3000\,\kms\ in][]{Tanikawa_2018}, though more studies will need to be done to test the robustness of this estimate.

Explosion mechanisms that would result in a greater amount of unburnt carbon in the ejecta include the pure deflagration \citep{Nomoto_1984b} or pulsating delayed detonation \citep{Hoeflich_1995,Dessart_2014} of a near-\Mch\ WD, and the violent merger of binary WDs \citep{Raskin_2014}. None of these models reproduce the unusual red color in the early light curves or the peculiar spectroscopic features of \sn\ as well as the helium-shell double-detonation scenario does.

We conclude that while it remains to be questioned if double detonations could really produce strong carbon features, the detection of \ion{C}{2} $\lambda\lambda$6580, 7234 lines in \sn\ does not necessarily invalidate our hypothesis of its double-detonation origin.

\section{Conclusions} \label{sec:conclusion}
We have presented observations of \sn, a peculiar SN\,Ia. \sn\ has an unusual color evolution, with a remarkably red $g_\mathrm{ZTF}-r_\mathrm{ZTF}$ color at early times due to continuous absorption in the blue portion of its SED. Absorption features observed around maximum light simultaneously suggest high (a blue continuum and shallow \ion{Si}{2} lines similar to those of overluminous, 99aa-/91T-like SNe) and low (the tentative \ion{Ti}{2} features resembling those of subluminous, 91bg-like SNe) photospheric temperatures. The nebular-phase spectra of \sn\ suggest a high ionization and low Ni abundance in the ejecta, consistent with a sub-\Mch\ explosion.

The early red colors are most likely due to a layer of IGEs in the outermost ejecta as products of a helium-shell detonation, in the sub-\Mch\ double-detonation scenario. If the asymmetric ejecta are observed from the hemisphere opposite to the helium ignition point, we find that the resultant synthetic spectra could qualitatively reproduce some of the observed properties, including (i) significant line-blanketing of flux due to IGEs at early phases; (ii) strong absorption features around 4200\,\r{A} as well as relatively weak \ion{Si}{2} features near maximum brightness, and (iii) blueshifted [\ion{Fe}{2}] $\lambda$7155 accompanied with a relatively low expansion velocity of \ion{Si}{2} at peak. Current double-detonation models cannot reproduce the strong \ion{C}{2} lines in \sn\ at its maximum luminosity, but the double-detonation hypothesis is not necessarily disfavored and an improved RT treatment will be needed in modeling the observational effects of unburnt carbon.
No existing double-detonation model can fully explain all the observational properties of \sn. As a result, it is possible that some alternative model is superior, though we find that the early red colors are difficult to explain with alternative explosion scenarios. Further refinement of multidimensional models covering a finer grid of progenitor properties may answer the question if the peculiar \sn\ is really triggered by a double detonation.\\

%\begin{acknowledgements}

{We thank the anonymous referee for a thoughtful report.} We are grateful to Ping Chen, Avishay Gal-Yam, Anthony Piro, Anna Ho, and Jiaxuan Li for fruitful discussion, as well as Peter Blanchard and Jillian Rastinejad for the LRIS spectra they obtained. We acknowledge U.C. Berkeley undergraduate students Kate Bostow, Cooper Jacobus, Gabrielle Stewart, Edgar Vidal, Victoria Brendel, Asia deGraw, Conner Jennings, and Michael May for the Lick/Nickel photometry. K.J.S. was in part supported by NASA/ESA Hubble Space Telescope programs \#15871 and \#15918. S.J.B. and D.M.T. acknowledge support from NASA grant HST-AR-16156. L.H. is funded by the Irish Research Council under grant GOIPG/2020/1387. K.M. is funded by the EU H2020 ERC grant 758638. S.S. acknowledges support from the G.R.E.A.T. research environment, funded by {\em Vetenskapsr\aa det},  the Swedish Research Council, project 2016-06012. G.D. is supported by the H2020 European Research Council grant 758638. C.D.K. is partly supported by a CIERA postdoctoral fellowship. A.V.F.'s supernova group at U.C. Berkeley received generous financial assistance from the Christopher R. Redlich Fund, Briggs and Kathleen Wood (T.G.B. is a Wood Specialist in Astronomy), Alan Eustace (W.Z. is a Eustace Specialist in Astronomy), and numerous other donors.

We appreciate the excellent assistance of the staffs at the observatories
where data were obtained. 
This work is based on observations obtained with the Samuel Oschin Telescope 48-inch and the 60-inch Telescope at the Palomar Observatory as part of the Zwicky Transient Facility project. ZTF is supported by the NSF under grant AST-2034437 and a collaboration including Caltech, IPAC, the Weizmann Institute of Science, the Oskar Klein Center at Stockholm University, the University of Maryland, Deutsches Elektronen-Synchrotron and Humboldt University, the TANGO Consortium of Taiwan, the University of Wisconsin at Milwaukee, Trinity College Dublin, Lawrence Livermore National Laboratories, IN2P3, University of Warwick, Ruhr University Bochum and Northwestern University. Operations are conducted by COO, IPAC, and UW.
The SED Machine is based upon work supported by the NSF under grant 1106171. The ZTF forced-photometry service was funded under the Heising-Simons Foundation grant \#12540303 (PI: Graham). The Gordon and Betty Moore Foundation, through both the Data-Driven Investigator Program and a dedicated grant, provided critical funding for SkyPortal.

A major upgrade of the Kast spectrograph on the Shane 3\,m telescope at Lick Observatory, led by Brad Holden, was made possible through generous gifts from the Heising-Simons Foundation, William and Marina Kast, and the University of California Observatories.  KAIT and its ongoing operation were made possible by donations from Sun Microsystems, Inc., the Hewlett-Packard Company, AutoScope Corporation, Lick Observatory, the U.S. NSF, the University of California, the Sylvia \& Jim Katzman Foundation, and the TABASGO Foundation. Research at Lick Observatory is partially supported by a generous gift from Google.

This work is also based in part on observations made with the Nordic Optical Telescope, owned in collaboration by the University of Turku and Aarhus University, and operated jointly by Aarhus University, the University of Turku, and the University of Oslo (respectively representing Denmark, Finland, and Norway), the University of Iceland, and Stockholm University at the Observatorio del Roque de los Muchachos, La Palma, Spain, of the Instituto de Astrofisica de Canarias. The W. M. Keck Observatory is operated as a scientific partnership among the California Institute of Technology, the University of California, and NASA; the observatory was made possible by the generous financial support of the W. M. Keck Foundation. Observations reported here were obtained at the MMT Observatory, a joint facility of the Smithsonian Institution and the University of Arizona. W. M. Keck Observatory and MMT Observatory access was supported by Northwestern University and the Center for Interdisciplinary Exploration and Research in Astrophysics (CIERA).

This work has made use of data from the Asteroid Terrestrial-impact Last Alert System (ATLAS) project, which is primarily funded to search for near-Earth objects (NEOs) through NASA grants NN12AR55G, 80NSSC18K0284, and 80NSSC18K1575; byproducts of the NEO search include images and catalogs from the survey area. This work was partially funded by Kepler/K2 grant J1944/80NSSC19K0112 and HST GO-15889, and STFC grants ST/T000198/1 and ST/S006109/1. The ATLAS science products have been made possible through the contributions of the University of Hawaii Institute for Astronomy, the Queen's University Belfast, the Space Telescope Science Institute, the South African Astronomical Observatory, and The Millennium Institute of Astrophysics (MAS), Chile.

%\end{acknowledgements}

\facility{PO:1.2m (ZTF), Swift (UVOT), KAIT, Nickel, Liverpool:2m (IO:O), PO:1.5m (SEDM), FTN (FLOYDS), FTS (FLOYDS), NOT (ALFOSC), Liverpool:2m (SPRAT), Keck:I (LRIS), MMT (Binospec), Hale (TSpec).}
\software{\texttt{astropy} \citep{Astropy_2013, Astropy_2018},
\texttt{dynesty} \citep{Speagle_dynesty_2020}, 
\texttt{matplotlib} \citep{Matplotlib_2007}, 
\texttt{NumPy} \citep{numpy_2020},
\texttt{prospector} \citep{Johnson_prospector_2021}, 
\texttt{PyMC} \citep{pymc_2016},
\texttt{PypeIt} \citep{pypeit:zenodo}, 
\texttt{pysedm} \citep{Rigault_pysedm_2019},
\texttt{sncosmo} \citep{Barbary_SNCosmo_2023},
\texttt{Python-FSPS} \citep{Conroy_2009,Conroy_2010,ForemanMackey_FSPS_2014},
\texttt{Sedona} \citep{Kasen_Sedona_2006}, 
\texttt{ZFPS} \citep{Masci_ZTFforced_2023}.
}

\bibliography{SN2022joj, software, telescope}
\bibliographystyle{aasjournal}



\end{CJK*}
\end{document}